# Two regimes of interaction of a Hot Jupiter's escaping atmosphere with the stellar wind and generation of energized atomic hydrogen corona

Shaikhislamov[1] I. F., Khodachenko[2,3] M. L., Lammer[2] H., Kislyakova[2] K. G., Fossati[2] L., Johnstone[4] C. P., Prokopov[1] P. A., Berezutsky[1] A. G., Zakharov[1] Yu.P., Posukh[1] V. G.

1) Institute of Laser Physics SB RAS, Novosibirsk, Russia
2) Space Research Institute, Austrian Acad. Sci., Graz, Austria
3) Skobeltsyn Institute of Nuclear Physics, Moscow State University, Moscow, Russia
4) Dep. of Astrophysics, University of Vienna

E-mail address: maxim.khodachenko@oeaw.ac.at

**Abstract:**

The interaction of escaping upper atmosphere of a hydrogen rich non-magnetized analog of HD209458b with a stellar wind of its host G-type star at different orbital distances is simulated with a 2D axisymmetric multi-fluid hydrodynamic model. A realistic sun-like spectrum of XUV radiation which ionizes and heats the planetary atmosphere, hydrogen photo-chemistry, as well as stellar-planetary tidal interaction are taken into account to generate self-consistently an atmospheric hydrodynamic outflow. Two different regimes of the planetary and stellar winds interaction have been modelled. These are: 1) the *"captured by the star"* regime, when the tidal force and pressure gradient drive the planetary material beyond the Roche lobe towards the star, and 2) the *"blown by the wind"* regime, when sufficiently strong stellar wind confines the escaping planetary atmosphere and channels it into the tail. The model simulates in details the hydrodynamic interaction between the planetary atoms, protons and the stellar wind, as well as the production of energetic neutral atoms (ENAs) around the planet due to charge-exchange between planetary atoms and stellar protons. The revealed location and shape of the ENA cloud either as a paraboloid shell between ionopause and bowshock (for the *"blown by the wind"* regime), or a turbulent layer at the contact boundary between the planetary stream and stellar wind (for the *"captured by the star"* regime) are of importance for the interpretation of Lyα absorption features in exoplanetary transit spectra and characterization of the plasma environments.

Keywords: hydrodynamics – plasmas – planets and satellites: individual: exoplanets – planets and satellites: physical evolution – planets and satellites: atmosphere – planet–star interactions

## 1. Introduction

The Hot Jupiter HD 209458b is probably one of the best studied exoplanets which is extensively used as a classical example of a close-orbit gas giant for the different types of modelling and concept prove. It is the first transiting exoplanet for which atmospheric absorption was observed in the resonance lines e.g. NaI (*Charbonneau et al. 2002*). The measurements of Lyα line absorption during transits (*Vidal-Madjar et al. 2003, Ben-Jaffel 2007, Vidal-Madjar et al. 2008; Ben-Jaffel 2008; Ehrenreich et al. 2008*) yielded a value of about 10% at velocity range ±150 km/s. That was the first indication of an extended atmosphere within the planet's Roche lobe which, according to the observed Doppler-shifted absorption in the lines of heavy species such as CII or OI, appears in a blowing off state. Spectral transit observations nowadays provide information about the composition and dynamics of material in the nearby vicinity of exoplanets. The measurements in

HI, as well as in the lines of other species, e.g., CII (*Linsky et al. 2010*), MgI (*Vidal-Madjar et al. 2013*) show more absorption in the blue wing of the line, indicating a larger tail-wards outflow, rather than particles motion towards the star. A definite blue-shifted absorption in Lyα observed, besides of HD209458b, also in the Hot Jupiter HD189733b (*Lecavelier des Etangs et al. 2010; Bourrier et al. 2013*) and for a colder giant 55 Cnc (*Ehrenreich et al., 2012*), suggests that these planets have expanding atomic H envelopes which undergo partial transits when the planets graze the stellar disc. The recent detection of a giant exosphere surrounding the warm Neptune GJ 436 b (*Kulow et al. 2014, Ehrenreich et al. 2015*) provides evidence that the neutral hydrogen atoms are not swept away, like in the case for the above mentioned evaporating hot Jupiters, but are dispersed within a large volume around the planet (*Bourrier et al. 2015, 2016*). The NUV transit observations of the giant WASP12b (*Fossati et al. 2010ab, Haswell et al. 2012*) reveal that similarly to HD209458b and HD189733b, this exoplanet also has an evaporating exosphere. Moreover, the detected broad depression of the normally bright emission cores in the Mg II resonance lines of the WASP12b host star may be interpreted as absorption by material ablated from the planet (*Haswell et al. 2012, Fossati et al. 2013*). Despite of consensus that Hot Jupiters like HD 209458b have an extended exosphere and are losing mass by hydrodynamic (HD) flow, several important details, due to many uncertainties, are not yet completely elucidated from observations. That concerns in particular the questions: whether the exosphere size is comparable to the Roche lobe or significantly larger; what are the outflow speed and mass loss rate; what is the spatial structure of the escaping planetary material flow.

The progress in observations of close-orbit giants stimulated development of theoretical and numerical models (e.g., *Lammer et al. 2003, Yelle 2004, Lecavelier des Etangs et al. 2004, Jaritz et al. 2005, Tian et al. 2005, Koskinen et al. 2007, Garcia Munoz 2007, Schneiter et al. 2007, Murray-Clay et al. 2009, Guo 2011, Trammel et al. 2011, Shaikhislamov et al. 2014, Khodachenko et al. 2015*). The predictions of these models are rather different due to different treatment of various factors, affecting the heating and expansion of a Hot Jupiter's upper atmosphere. Among such factors are the amount of radiative energy input and its spatial distribution; material cooling due to expansion and radiation; the effects of tidal force, radiation pressure, ionization and recombination, hydrogen chemistry, the presence of metals in the outflow, as well as boundary conditions at the planet surface. In spite of certain quantitative difference, the qualitative outcome of all the numerical models applied to the Hot Jupiter HD 209458b is that it generates an escaping super-sonic hydrodynamic planetary wind (PW), with a temperature of about $10^4$ K and velocity of the order of 10 km/s, which results in the mass loss rates of about $\sim 10^{10}$–$10^{11}$ g s$^{-1}$. It has been also shown recently with a dedicated numerical modelling (*Khodachenko et al. 2015*) that a hypothetic planetary magnetic field with values above ~0.3 G should affect the PW structure and significantly reduce the mass loss.

However, the application of the existing models to observations still remains deficient. While the simulated density of an extended atmosphere of a Hot Jupiter is high enough to provide the column density needed for significant absorption, the predicted expansion speed of material cannot explain the observed spectral features at velocities of several tens and up to ±150 km/s for hydrogen, or ±50 km/s for CII (*Linsky et al. 2010*). By analyzing the available data on HD 209458b transit, *Ben-Jaffel and Sona Hosseini* (*2010*) argue that the observed absorption features in HI can be due to either hydrogen atoms with temperature $\sim 10^4$ K filling the Roche lobe of the planet and providing sufficiently large column density of about $10^{21}$ cm$^{-2}$, or due to much faster atoms with thermal or direct velocities of the order of 100 km/s, resulting in much lower column density of about $3 \cdot 10^{13}$ cm$^{-2}$, but extended beyond the Roche lobe. At the same time, the data on the absorption in CII and OI can be explained only by the presence of fast particles related either with regular flows with velocities of at least 20 km/s, or with high temperatures of order of $10^6$ K.

To explain the above mentioned effects, the acceleration of slow hydrogen atoms by stellar radiation to high velocities was proposed in *Vidal-Madjar et al.* (*2003*). As it was shown in the case of HD209458b (*Lecavelier des Etangs et al. 2008, Bourrier & Lecavelier des Etangs 2013*), the radiation pressure is sufficiently strong to drive the planetary hydrogen atoms and to produce the absorption in the blue wing of Lyα. Another way to explain the presence of neutral hydrogen with even higher velocities, put forward by *Holmstrom et al.* (*2008*), involves the charge exchange between fast protons of the stellar wind (SW) and slow neutral hydrogen of the escaping PW. The created in such a way energetic neutral atoms (ENAs) are well known in the solar system, e.g. Earth (Collier et al. 2001), Mars (Futaana et al. 2006) and Venus (Galli et al. 2008). Calculations of Lyα absorption by ENAs in (*Holmstrom et al. 2008, Tremblin & Chiang 2013, Kislyakova et al. 2014a,b, Christie et al. 2016*) can to certain extent explain the mentioned above high velocity spectral features measured on HD209458b. Modeling of ENA clouds with 3D particle codes that include both radiation pressure and charge-exchange was used to interpret velocities of neutral hydrogen as high as 200 km/s observed in HD189733b (*Bourrier & Lecavelier des Etangs 2013*) and to put constraints on magnetic field (*Kislyakova et al. 2014a*) by distinguishing between different shapes and positions of ionopause or magnetopause from which planetary neutrals carried by PW are supposed to be injected into SW. In that respect, a combination of HD/MHD models of PW and SW interaction with 3D particle codes that operate in collisionless region further away from the planet opens a way for self-consistent modelling of the Hot Jupiter environments, which combines the advantages of both HD/MHD and particle codes and eliminates deficiencies of their separate use.

*Murray-Clay et al.* (*2009*) estimated that for HD 209458b the typical SW ram pressure equals the PW thermal and ram pressures close to the point where the escaping atmospheric flow becomes a super-sonic one. This conclusion was confirmed later in *Shaikhislamov et al.* (*2014*) and *Khodachenko et al.* (*2015*) who estimated the location of the pressure balance point at distances of about $5R_p$, which are in fact beyond the Roche lobe. Despite of the obvious need to complement the above mentioned hydrodynamic and particle models, widely used for the interpretation of observations, there have been so far only a few attempts of detailed study of the problem of interaction of SWs with the expanding atmospheres of Hot Jupiters. The reasons for that are several. First, the problem of PW and SW interaction is at least 2D, and therefore, it requires a more complicated numerical model as compared to 1D models used quite often. Second, at least two widely different fluids, i.e. those of the stellar and of the planetary origin, should be modeled. Third, the adequate modelling requires combining of the global interaction of the stellar and planetary plasma flows with the micro-physics of the PW generation. There is a number of works where first two of the above mentioned difficulties have been resolved by exploiting the available generic MHD or astrophysical codes adopted for the considered particular problem. For example, one of the first simulations of this type by *Stone and Proga* (*2009*) model the formation of a bowshock under the conditions of colliding winds. At the same time, this modelling still skips several important processes, such as self-consistent physical drive of the PW, tidal force, adiabatic cooling etc. *Tremblin & Chiang* (*2013*) put forward an important idea that to simulate the generation of ENAs, four hydrogen fluids should be considered, i.e. those attributed to the ionized and neutral components of the planetary and stellar origin. However, their model contains several significant simplifications. In particular, besides of assuming the escaping PW to originate at a prescribed distance of 4 planetary radii, the model ignores the XUV and electron impact ionization as well as the tidal forces. The main conclusion made in *Tremblin & Chiang* (*2013*) is that ENAs are generated in a mixing contact layer of the PW and SW fluids which is affected by Kelvin-Helmholtz instability and turbulence. To the same conclusions on the turbulent mixing character in the contact layer came *Christie et al.* (*2016*), who employed instead of a planar 2D geometry, a more realistic 2D axisymmetric one. Further simulations, done with astrophysical codes, revealed a complex 3D structure of the PW flow in the surrounding SW. For extremely close-in planets like Wasp-12b, the planetary material flow is pulled by stellar gravity, twisted by planet rotation, disrupted by interaction with the stellar wind, and spirals towards the star in irregular clumps (*Bisikalo et al 2013*). As summarized by *Matsakos et al.* (*2015*) on the basis of a parametric study, if the PW is stopped by SW inside of the Roche lobe, then a comet like structure is formed with a trailing tail. In opposite case a stream of escaping material is formed on the dayside as well that falls towards the star by spiraling just for a quarter of an orbit. The absence of accretion disk in that case is explained by interaction of the planetary escaping material with the SW which slows down and disrupts the PW stream removing significant portion of the angular momentum.

However, none of the above cited works, based on either particle or HD/MHD models, properly addresses the problem of consistent combination of the global planetary and stellar plasmas interaction with micro-physics of the PW generation. The widely used approach there consists in launching of PW by adopting a suitable boundary condition, instead of self-consistent account of its generation due to XUV heating. A spherically symmetric isothermal approximation, which is a too oversimplified assumption for PW, is considered quite often. It turns the PW into a super-sonic flow, which in reality, due to the presence of a SW barrier, can be well a sub-sonic one. Another severe limitation of a number of previous studies is the absence of the tidal force. The importance of the last has been demonstrated in 3D astrophysical simulations which, at the same time, did not include the neutral components of the flow at all. Having in mind the open questions and oversimplifications of the available nowadays models, we improve in the present paper the approach of *Tremblin & Chiang* (*2013*) by implementing a self-consistent 2D axisymmetric model which incorporates all important processes and factors regarding generation of the PW and is free of any artificial substitutions and prescriptions. To avoid any local balance approximations, the model simulates the dynamics of four interrelated fluids, i.e. the ionized and neutral components of the PW and SW, respectively. The self-consistently calculated neutral fraction of the PW in the region where it comes in contact with SW in our simulations appears significantly lower than the presumed value ¼ taken by *Tremblin & Chiang* (*2013*) based on work of *Murray-Clay et al.* (*2009*). The reason is that in calculation of free-expanding PW (i.e. without account of the braking effect of the incoming SW), its velocity is significantly higher, and for a particular distance the time-integrated probability of photo-ionization is smaller than in case when the SW forms a barrier.

Another crucial difference of the performed modelling concerns the circumstances of ENA production in the colliding PW and SW plasma flows. According to *Tremblin & Chiang* (*2013*) and *Christie et al.* (*2016*), all kinds of planetary atmospheric particles (i.e. ionized and neutral components of the PW), due to strong collisional coupling, are supposed to move with the same velocity, i.e. as a single fluid, and the penetration of atoms into SW is possible only within a kind of an interface mixing layer between the planetary and stellar proton flows. Reproduction of the mixing, that involves turbulent cascade from large to progressively smaller eddies cannot be reliably done without implementing of specific numerical algorithms (*Chernyshov et al. 2006, Miesch et al. 2015*). A good example here is Earth's magnetosphere boundary layer (magnetosheath) whose width is turbulently dynamic and varies from several up to 10 ion gyro-radii. We pay attention to the fact that at such an interface layer the planetary atoms decouple from protons and proceed further into the rarified SW or the shocked region, where the charge-exchange length is relatively large. The ground for such decoupling is the fact that the expanding PW protons are stopped by the SW protons, whereas PW atoms are not. The coupling between ionized species of the PW and SW takes place through the chaotic or regular magnetic fields, which are always present in the flows, rather than due to Coulomb collisions. Therefore, the mixing of planetary atoms with SW is laminar and much more effective and independent of planetary protons and turbulent interface mixing. In view of that, the underlying mechanism of ENAs production in our model is principally different from that considered in *Tremblin & Chiang* (*2013*) and is similar to a simple Monter-Carlo approach in which planetary atoms are injected from the prescribed boundary into the collisionless SW (*Holmstrom et al. 2008, Kislyakova et al. 2014a,b*).

Another important factor is the tidal force which causes two main regimes of interaction between PW and SW (*Matsakos et al. 2015*): 1) the so called *"captured by the star"* regime, when the tidal force and pressure gradient lead to the streaming of PW towards the star, and 2) the *"blown by the wind"* regime when sufficiently strong SW confines the escaping planetary atmosphere material and channels it away from the star as a kind of tail. In the present paper we simulate the interaction between the SW and escaping PW taking HD 209458b as a prototype of a Hot Jupiter, i.e. considering an exoplanet with the same as for the HD 209458b mass $M_p = 0.71M_J$ and radius of $R_p =1.38R_J$ (scaled in the solar system Jupiter's mass $M_J= 1.89 \times 10^{27}$ kg and radius $R_J= 71.5 \times 10^3$ km), orbiting around a solar-type G-star with the mass of $M_{st} = 1.148M_{Sun}$ and the age of ~4 Gyr (The Extrasolar Planets Encyclopaedia: http://www.exoplanet.eu). The conditions of a slow and fast stellar wind of a Sun-like star at different orbital distances D of the planet are considered. By this, it turns out, that for the real HD209458b (i.e. at D=0.045AU) both regimes of the PW and SW interaction, which are very different by their potentially observational features, could be realized simply due to insignificant (by several times) natural variations of the SW pressure. The quantitative characterization of atomic hydrogen cloud for the mentioned above two regimes of HD209458b atmosphere blow-off is one of the main goals of the present paper. At the same time, the proposed 2D model does not cover the essentially 3D effect of rotational twisting of the escaping PW stream. The latter might be relevant for the study of global large scale distribution of the lost planetary material inside the astrosphere, but at the same time, it is less important in the context of total planetary mass loss, as well as for the structure of the relatively close to the planet PW and SW interaction region and the related ENA cloud formation.

The main finding of the present work concerns the structure of the ENA cloud around the planet and its density which appears to be quite low. Due to photo-ionization of hydrogen, which for the solar type radiation at the distance of 0.045 AU has typical time of about 4 hours, and relatively low sub-sonic velocity of planetary flow affected by SW, the number of atoms reaching ionopause of HD209458b at about $5R_p$, is rather small, and the typical ENAs column density is about $10^{11}$ cm$^2$. This puts certain constraints on the interpretation, within the existing models, of the detected for this exoplanet, Lyα transit spectra as the result of absorption by ENAs and might require further detailed treatment of the involved physical concepts and approaches.

The paper is organized in the following way. In Section 2 we describe the modelling concept paying attention to the basic processes, equations, and geometry assumptions. In Section 3 we provide a qualitative picture of the PW and SW interaction, and discuss the particulars for the formation of ENA cloud around a planet. In Section 4 the results of numerical simulations are presented. Section 5 is dedicated to the discussion of the modeling results and conclusions.

## 2. Modelling concept

### 2.1 *Basic processes, equations and assumptions*

To simulate the interaction of planetary and stellar winds, we apply a multi-fluid model which includes the following species: $H, H^+, H_2, H_2^+, H_3^+$ of the planetary origin and $H, H^+$ of the stellar origin. A pure molecular hydrogen atmosphere of an HD209458b analogue in a barometric equilibrium with the base (inner boundary) temperature of 1000 K is taken as the initial state for simulation of the expanding PW driven by the stellar XUV radiative heating. The PW plasma is treated as a quasi-neutral fluid in thermal equilibrium with $T_e = T_i$. Note, that the neutral hydrogen component is not originally present in the SW. It is produced by charge exchange between the stellar protons and planetary atoms. The applied model is an extension of our model of Hot Jupiter's PW presented in *Shaikhislamov et al.* (*2014*) and *Khodachenko et al.* (*2015*). The new element is the inclusion of an SW and separate treatment of all species. As it was shown in our previous works

(*Shaikhislamov et al. 2014*, *Khodachenko et al. 2015*), as well as in *Yelle* (*2004*) and *Koskinen et al.* (*2007*), the components $H_2, H_2^+, H_3^+$ can exist around an HD 209458b analogue planet only at very low heights $< 0.1 R_p$ in a relatively dense thermosphere. Thus, the interaction between the PW and SW, which takes place at higher altitudes of several planetary radii $R_p$, involves only protons and hydrogen atoms. To distinguish between the species of planetary and stellar origin, we use for the corresponding physical parameters an upper subscript *pw* and *sw* for the PW and SW, respectively. The lower subscripts are used to indicate a sort of particles, e.g., ionized "H+" and neutral "H" hydrogen. The final result of the modelling which will be presented here concerns the interaction of four major fluids – planetary protons $n_{H+}^{pw}$ and atoms $n_H^{pw}$, stellar protons $n_{H+}^{sw}$, which constitute the SW, and energetic neutral atoms (ENAs) $n_H^{sw}$ formed due to charge exchange between the SW protons and planetary neutrals. However, during the modelling calculation itself we retain also the molecular hydrogen species. The reason for inclusion of the hydrogen chemistry, besides of the need to account the IR cooling produced by $H_3^+$ molecule, is that a significant part of stellar XUV energy is absorbed at heights where molecular hydrogen dominates, as compared to the atomic one. In total, the inclusion of molecular hydrogen results in overall reduction of mass loss rate by a factor of 1.5. The model code solves numerically the continuity, momentum and energy equations, which can be written in the following form:

$$\frac{\partial}{\partial t} n_j + \nabla (\mathbf{V}_j n_j) = N_{XUV,j} + N_{exh,j} \quad (1)$$

$$m \frac{\partial}{\partial t} \mathbf{V}_j + m(\mathbf{V}_j \nabla)\mathbf{V}_j = -\frac{1}{n_j}\nabla n_j kT_j - \frac{z_j}{n_e}\nabla n_e kT_e - m\nabla U - m\sum_i C_{ji}^\upsilon (\mathbf{V}_j - \mathbf{V}_i) + m\eta_j \nabla^2 \mathbf{V}_j \quad (2)$$

$$\frac{\partial}{\partial t} T_j + (\mathbf{V}_j \nabla) T_j + (\gamma - 1) T_j \nabla \mathbf{V}_j = W_{XUV,j} - \sum_i C_{ji}^T (T_j - T_i) \quad (3)$$

As in our previous models (*Shaikhislamov et al. 2014*, *Khodachenko et al. 2015*), the main processes, responsible for the transformation between neutral and ionized hydrogen particles of the planetary origin are photo-ionization, electron impact ionization and dielectronic recombination. These are included in the term $N_{XUV,j}$ in continuity equation (1) and are applied for all species. Note, that the recombination is negligible for stellar particles due to high temperature of the SW. For the same reason the electron impact ionization is important only in SW where it exceeds the photo-ionization rate. Photo-ionization also results in a strong heating of planetary material by the produced photo-electrons (for SW the XUV heating effect is negligible). The corresponding term $W_{XUV,j}$ in the energy equation (3) (addressed in *Shaikhislamov et al. 2014*, *Khodachenko et al. 2015*) is derived by integration of the stellar XUV spectrum. For the solar type host star of HD209458b we use here as a proxy the spectrum of the Sun, compiled by (*Tobiska 1993*) and covering the range 10–912 Å, binned by 1 Å. The spectrum is based on measurements of solar radiation under moderate activity conditions with proxy index $P_{10.7} = 148$. Total integrated flux at 1 AU in this model is 4.466 erg s$^{-1}$ cm$^{-2}$. We note that it is only by a factor of 1.5 smaller than the XUV flux of HD209458 reconstructed from the measurements in UV and X-ray bands (*Louden et al. 2016*). This restricts the similarity of the modelled stellar-planetary system to the real HD209458 system, when other parameters of both systems are considered to be the same. It is assumed that the energy released in the form of photo-electrons is rapidly and equally re-distributed between all locally present particles with efficiency of $\eta_h = 0.5$. This is a commonly used assumption, which we adopted on the basis of qualitative analysis (*Shaikhislamov et al. 2014*).

Another kind of important interaction between the considered particle populations is resonant charge-exchange collisions. Indeed, charge-exchange has the cross-section of about $\sigma_{exc} = 6 \cdot 10^{-15} \, cm^2$ at low energies, which is an order of magnitude larger than the elastic collision cross-section. Experimental data on the differential cross-sections can be found, for example, in *Lindsay & Stebbings (2005)*. When planetary atoms and protons slip relative each other, because they have different thermal pressure profiles and protons feel electron pressure while atoms do not, the charge-exchange between them leads to velocity and temperature interchange. We describe this process with a collision rate $C_{ji}^{\upsilon}$ where upper index indicate the value being interchanged. For example, in momentum equation for planetary protons it looks as $C_{H^+H}^{\upsilon} = n_H^{pw}\sigma_{exh}\upsilon$, where the interaction velocity $\upsilon \approx \sqrt{V_{Ti}^2 + V_{Tj}^2 + (V_j - V_i)^2}$ depends in general on thermal and relative velocities of the interacting fluids (in the considered example, protons and neutral atoms). More precise expressions for charge-exchange terms in the continuity, momentum and energy equations, obtained by averaging of the collision operator over the Maxwell distribution (e.g., *Meier & Shumlak 2012*), differ from those used in our work by an order of unity coefficients or by small additional terms, which are inessential for the present study.

The cross-section of resonant charge-exchange decreases in a certain way with the increasing interaction velocity $\upsilon$. Therefore, in the present paper we work directly with the product $\sigma_{exh}\upsilon$ obtained according to experimental data in *Lindsay & Stebbings (2005)*. For example, for sufficiently cold planetary particles with relative $\upsilon \approx 10^6 \, cm/s$ the charge-exchange rate in the PW is equal to $\sigma_{exh}(\upsilon) \cdot \upsilon \approx 6 \cdot 10^{-9} \, cm^3/s$ and it is much smaller than that in the much faster stellar wind ($\approx 6 \cdot 10^{-8} \, cm^3/s$, for $\upsilon = V_{H+}^{SW,slow} = 3 \cdot 10^7 \, cm/s$ and $\approx 8 \cdot 10^{-8} \, cm^3/s$ for $\upsilon = V_{H+}^{SW,fast} = 6 \cdot 10^7 \, cm/s$). For the densities above $10^5 \, cm^{-3}$ and other parameters typical for HD209458b, the charge-exchange between planetary atoms and protons ensures that they move practically together with a relatively small slippage and equal temperatures, constituting, therefore, a hydrodynamic PW in thermal equilibrium. As to the SW, the charge-exchange between stellar protons and atoms is rather rare, given that the typical mean-free path is of an order of $10 \, R_p$ (i.e. $> 10^{11} \, cm$).

There is also charge-exchange between planetary and stellar particles which is responsible for the production of ENAs. Within the frame of our multi-species model this process is described as a double exchange (*Tremblin & Chiang 2013*). When a slow planetary atom undergoes charge-exchange with a fast stellar proton, a slow planetary proton and a fast stellar atom (ENA) are produced. Since the velocity and temperature of planetary protons and atoms are close to each other, and the same is true for the SW species, this process can be described as an act of interchange between planetary particles on the one hand and stellar particles on the other. The account of this process of ENAs formation requires a multi-species approach, because if described without distinguishing between the origin of particular particle populations e.g., planetary and stellar, the charge-exchange between particles would lead to the formation of essentially different (in velocity and temperature) sub-populations, which cannot be described within a single particle sort. The volume rate of double exchange is given by $N_{exh} = n_H^{pw} n_{H+}^{sw} \sigma_{exh}\upsilon - n_{H+}^{pw} n_H^{sw} \sigma_{exh}\upsilon$. In this case the interaction velocity in most of cases is practically equal to the SW velocity, because it is much higher than the particles thermal speed. However, in the shocked region of the colliding planetary and stellar winds, the thermal velocity becomes dominant.

Besides of charge-exchange we also take into account elastic collisions. First of all, the interaction of PW and SW is mediated in our model *by effective collisions between planetary and stellar protons*. In reality the interpenetration of counter-streaming plasma flows is restricted by gyrorotation in a background magnetic field. Frozen-in magnetic fields, either laminar or chaotic, are always present in both stellar and planetary plasmas. Because of large characteristic scale of the considered system $\sim 10^{10}$ cm the proton gyroradius remains (comparatively) very small. Even for a very weak magnetic field of 1 nT it is still at least ten times smaller. Thus, for expected sporadic magnetic fields of the order of $10^3$ nT in the stellar wind at the considered orbital distances (~0.47 AU), the interpenetration of planetary and stellar protons should be microscopically small. In this case the protons interaction can be safely described by effective collisions with a sufficiently small mean free path (of about a gyroradius), which at the density of $n_{H+}^{pw} \sim 10^6$ cm$^{-3}$ corresponds to a cross section of the order of $\sigma_{H^+H^+,eff} \sim 10^{-12}$ cm$^2$. We note that Coulomb collisions are negligible here because of high energy of stellar protons. In addition to the proton-proton coupling we include planetary and stellar atomic-atomic and proton-atomic elastic collisions with the cross-sections of $\sigma_{HH} \approx \sigma_{H+H} \approx 10^{-16}$ cm$^2$ which, however, are relatively unimportant.

|  | planetary protons | planetary atoms | stellar protons | stellar atoms |
|---|---|---|---|---|
| $N_{XUV,j}$ | $N_1 = n_{PH}^{pw} \langle \sigma_{XUV} F_{XUV} \rangle - n_{H+}^{pw} n_e V_{Te} \sigma_{rec}$ | $-N_1$ | $N_2 = n_H^{sw} \langle \sigma_{XUV} F_{XUV} \rangle$ | $-N_2$ |
| $N_{exh,j}$ | $N_{exh} = n_H^{pw} n_{H+}^{sw} \sigma_{exh} \upsilon - n_{H+}^{pw} n_H^{sw} \sigma_{exh} \upsilon$ | $-N_{exh}$ | $-N_{exh}$ | $N_{exh}$ |
| $C_{ji}^\upsilon$ | $C_{pw}^\upsilon = n_{H+}^{pw} n_H^{pw} \sigma_{exh} \upsilon$ | $C_{pw}^\upsilon$ | $C_{sw}^\upsilon = n_{H+}^{sw} n_h^{sw} \sigma_{exh} \upsilon$ | $C_{sw}^\upsilon$ |
| $C_{H^+H^+}^\upsilon, C_{HH}^\upsilon$ | $C_{H^+H^+}^\upsilon = n_{H+}^{pw} n_{H+}^{sw} \sigma_{H^+H^+,eff} \upsilon$ | $C_{HH}^\upsilon = n_H^{pw} n_H^{sw} \sigma_{HH} \upsilon$ | $C_{H^+H^+}^\upsilon$ | $C_{HH}^\upsilon$ |
| $C_{ji}^T$ | $C_{pw}^T = C_{pw}^\upsilon$ | $C_{pw}^T$ | $C_{sw}^T = C_{sw}^\upsilon$ | $C_{sw}^T$ |
| $C_{exh}^\upsilon$, $C_{exh}^T = C_{exh}^\upsilon$ | $N_{exh}/n_{H+}^{pw}$ | $N_{exh}/n_H^{pw}$ | $N_{exh}/n_{H+}^{sw}$ | $N_{exh}/n_H^{sw}$ |

**Table 1.** The list of exchange and collisional terms included in the model equations.

The momentum transfer between similar particles due to thermal motion in the fluid approach is described by viscosity. Its dimensionless value can be estimated as relation of mean-free path to the system size. For the planetary protons experiencing the Coulomb collisions at a temperature of $10^4$ K and density of $10^6$ cm$^{-3}$ the dimensionless viscosity is about $\eta_{H+}^{pw} \sim 10^{-5}$. This value is used in the simulations. In fact it practically doesn't influence the solution, but helps to improve the numerical stability. At the same time, no viscosity is considered in the SW, because it is really negligible. The *Table 1* lists in the explicit form the exchange and collisional terms appeared in equations (1)-(3) for the interacting PW and SW species.

Other (heavier) species included in the PW model, such as $H_2$, $H_2^+$, $H_3^+$, exist only very close to the conventional planetary surface r=R$_p$ (the inner boundary of the simulation box) in the dense and strongly collisional thermosphere where all particles move together and specific collision cross-sections are not relevant. The dynamics of each of the involved molecular components of PW is described by an additional term $\sum_i R_{ji}$ in the right hand side of the continuity equation (1). Here $R_{ji}$ is a set of relevant reactions of hydrogen chemistry, described in *Khodachenko et al.* (*2015*) and the subscript *j* denotes the corresponding molecular hydrogen components i.e. $H_2$, $H_2^+$, $H_3^+$. As the molecular components exist only in a highly collisional layer of thermosphere, the approximation of equal velocities for all neutral and ionized components is valid, so that the dynamics of all ionized species as well as all neutral components of the PW is described by the corresponding combined momentum equations (2) taking into account the average particle mass of the neutral and ionized fluids:

$$\overline{m}_a = m \cdot (n_H + 2n_{H2})/n_H^{pw}, \quad \overline{m}_p = m \cdot (n_{H+} + 2n_{H2+} + 3n_{H3+})/n_{H+}^{pw} \quad (5)$$

The fluid velocities $\mathbf{V}_{Hj}^{pw}$ in these equations stay for $H, H_2$, and $H^+, H_2^+, H_3^+$, respectively.

Finally, the not listed in the *Table 1*, heating term for the PW particles is given by expression:

$$W_{XUV,j} = (\gamma - 1) \frac{\eta_h (n_H^{pw} + 2n_{H2}^{pw}) \cdot \langle (\hbar\nu - E_{ion})\sigma_{XUV} F_{XUV} \rangle - C_{IR} \cdot n_{H3+} - n_H^{pw} n_e \upsilon_{Te}(E_{21}\sigma_{12} + E_{ion}\sigma_{ion})}{n_H^{pw} + n_{H2}^{pw} + n_{H+}^{pw} + n_{H2+}^{pw} + n_{H3+}^{pw} + n_e} \quad (6)$$

Attenuation of XUV flux $F_{XUV}(\nu)$ is calculated for each spectral bin according to $\sigma_{XUV}(\nu)$. The photo-ionization cross-section at high energies is about twice larger for $H_2$ that for H. For the IR cooling rate $C_{IR}$ (due to $H_3^+$) we use in our model the one obtained in *Miller et al.* (*2013*). The last term in the numerator of equation (6) is the energy loss by electron impact excitation (Lyα cooling) and ionization. Note that we calculate the radiative cooling in a thin optical approximation. The reabsorption and transfer of Lyα photons, studied in our previous paper (*Shaikhislamov et al. 2014*), may increase the maximum temperature of a stationary planetary thermosphere (for HD 209458b analogue) by about 1500 K. However, due to zonal flows and adiabatic cooling this effect is likely to be inessential and therefore, it is not included in the present model.

## 2.2 *Geometry of the numerical model*

In the present study of the PW and SW interaction we employ a 2D axially symmetric hydrodynamic numerical model with the symmetry axis taken along the planet-star line in the reference frame of the tidally locked planet, which rotates synchronous with its orbital revolution (see *Figure 1*). Such 2D axially symmetric geometry is well suited for the simulation of PWs of tidally locked non-magnetized planets, illuminated by a star just from only one side (dayside). It is much more relevant than the 2D plane geometry used for a similar case in *Tremblin & Chiang* (*2013*). The relevance and limitations of the applied *quasi-axisymmetric* approximation were studied in details in *Khodachenko et al. (2015)*. In particular, within this approximation we disregard the Coriolis force and circularly average the centrifugal force. Such simplification is possible for sufficiently slowly rotating tidally locked planets. Note, that the planetary surface rotation velocity, which appears a symmetry breaking factor, is usually less than the considered typical velocities. For example for the considered tidally locked analogue of HD209458b at orbital distance of D = 0.045 AU the surface rotation velocity is about 3 km/s. It is certainly less than the typical PW outflow velocity of ~10 km/s. Thus, the Coriolis force is not expected to play significant role. Moreover, it doesn't change the energy of the moving material and its influence of the thermal mass loss rate is negligible. Another factor which breaks axial symmetry on the tidally locked planet is the centrifugal force. It has maximum value in the ecliptic plane and turns to zero on the rotation axis. The applied circular averaging around the planet-star line (as indicated in Figure 1) disregards this difference, which is also not crucial for the considered processes.

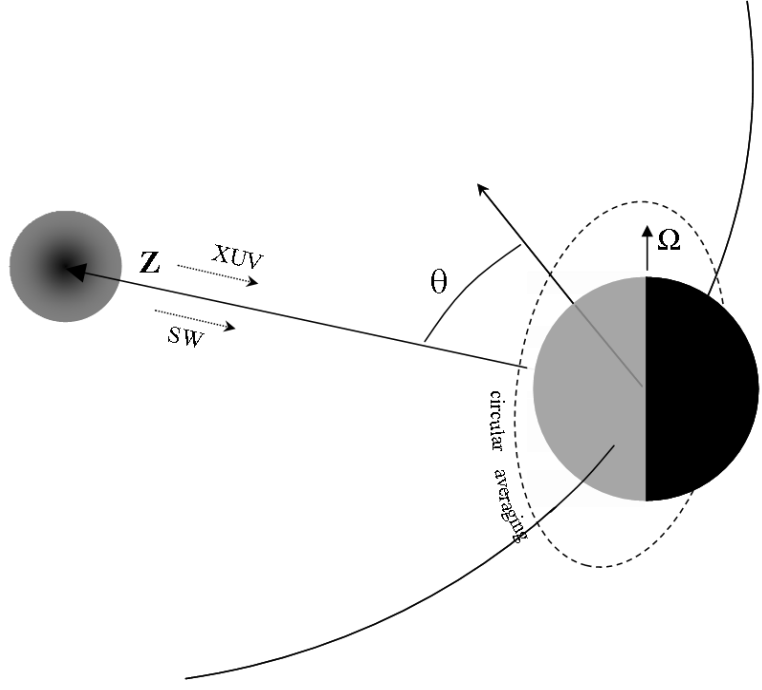

**Figure. 1** The geometry of numerical model in the frame of the planet-based spherical coordinate system $(R,\theta,\varphi)$ with the star located at $\theta=0$. *Quasi-axisymmetric* approximation requires circular averaging of the Coriolis and centrifugal forces as indicated by dash line around the planet-star axis. For the simulations a cylindrical coordinate system $(r = R\sin\theta, Z, \varphi)$ is used with Z-axis pointed towards the star.

Directly related with the geometry approximations is an expression for the gravitational stellar-planetary interaction which includes also the rotational effects. Averaged around the planet-star axis it reads for a tidally locked system (*Khodachenko at al. 2015*) in the planet-based spherical coordinate system $(R,\theta,\varphi)$ with the star located at $\theta=0$ as follows:

$$U = -\frac{GM_p}{R}\cdot\left(1+\frac{1}{2}\frac{R^3}{R_{L1}^3}\frac{7\cos^2\theta-1}{6}\right) \qquad (7)$$

Here $R_{L1} = D(M_p/3M_*)^{1/3}$ is the first Lagrange point.

One more limitation of the applied *quasi-axisymmetric* approximation for the tidally locked stellar-planetary system is the assumption of co-directionality of the relative velocity of SW flow and the ionizing and heating stellar radiation flux. In reality, the total velocity $\tilde{V}_{SW} = \sqrt{V_{SW}^2 + V_K^2}$ of an incoming SW, besides of the velocity $V_{SW}$ of the wind itself, should include also the Keplerian component of the planetary orbital motion $V_K = \sqrt{GM_{st}/D}$, which for a Sun-like star is $\approx 30/\sqrt{D}$ km/s, where D is expressed in AU. Note that for extremely close exoplanets, the orbital component $V_K$ becomes even a dominating part of the SW velocity vector $\tilde{V}_{SW}$, and in this case different directions of the SW the stellar radiation flux break the symmetry of the problem.

The velocity $V_{SW}$ of a Sun-type SW can be roughly approximated by a linearly increasing function below 0.1 AU and an asymptotic constant value $V_0$ beyond that, which may be taken equal to 350 km/s and 700 km/s for the slow and fast wind, respectively. Therefore, the ratio of the ram pressures caused by the SW itself and by the planetary orbital motion is $V_{SW}^2/V_K^2 = (V_0/3)^2 D^3$, i.e. $\sim 13{,}6 \times 10^3 D^3$ and $\sim 54{,}4 \times 10^3 D^3$ for the slow and fast SW, respectively. As the closest orbital distance considered in our work is that of HD209458b (i.e., D=0.045AU), the neglecting of orbital velocity can be justified for the fast SW. For the slow SW, the model can give at D=0.045AU only qualitative results, as the unaccounted ram pressure component due to the orbital motion is comparable to the SW ram pressure. However, at the distance of D=0.075 AU and beyond, the co-directional approximation for the total relative SW velocity and radiation flux along the planet-star line becomes sufficiently valid for both the slow and fast SW.

*2.3 Stellar Wind parameters*

We study the interaction of PW and SW for the case of an HD209458b analog exoplanet orbiting a Sun-like star within the distance range $0.045 AU \leq D \leq 0.12 AU$, which covers the observed orbits of known Hot Jupiters, as well as admits a possibility of an orbit change due to possible migration during a planet's life. In all considered cases, SW is assumed to be super-Alfvénic (relative velocity) and high-beta, so that the effects of the background magnetic field pressure remain insignificant, as compared to those of the plasma ram and thermal pressures, and both the PW, as well as the SW, could be treated as the non-magnetized ones. In fact, the orbital distance, D, is one of the most important parameters of the model, because several major factors which drive the system behavior, such as the XUV flux, the stellar wind ram pressure, and the tidal force depend on it strongly.

At the considered distances, the SW is still under acceleration. In our simulations we use the SW modelling data from *Johnstone et al. (2015)* which take this into account and reproduce the values for the wind speed as well as the density and temperature in the whole range of distances. Our modelling calculations are performed for two sets of the SW parameters, shown in the *Table 2*, which correspond to the slow and the fast SW, respectively. In general, fast SW pressure is about 1.5 times larger than that of the slow SW. No magnetic field is explicitly considered in our simulations.

| Orbital distance D [AU] | Orbital velocity $V_K$, [km/s] | Slow SW | | | Fast SW | | |
|---|---|---|---|---|---|---|---|
| | | $n_{sw}$ [cm$^{-3}$] | $V_{sw}$ [km/s] | $T_{sw}$ [×10$^6$ K] | $n_{sw}$ [cm$^{-3}$] | $V_{sw}$ [km/s] | $T_{sw}$ [×10$^6$ K] |
| 0.045 | 140 | 3430 | 230 | 1.30 | 1130 | 520 | 2.90 |
| 0.06 | 120 | 1690 | 260 | 1.23 | 620 | 570 | 2.80 |
| 0.075 | 110 | 990 | 290 | 1.19 | 360 | 600 | 2.70 |
| 0.09 | 100 | 645 | 310 | 1.10 | 240 | 620 | 2.60 |
| 0.12 | 87 | 355 | 340 | 0.90 | 130 | 660 | 2.00 |

**Table 2.** Parameters of slow and fast SW for different orbital distances D (according to *Johnstone et al. (2015)*), used in simulations.

**3. Qualitative picture of PW and SW interaction**

The PW is generated by stellar XUV heating. At low altitudes in the inner regions, it is mostly a neutral flow, but at distances of about $(3 \div 4) R_p$, PW becomes essentially photo-ionized, though not fully. Thermal and bulk velocities of such PW could reach up to 10 km/s (*Yelle 2004, Shaikhislamov et al. 2014*).

An important feature of the PW flow is rather high degree of collisional coupling, even relatively far from the planet, characterized by Knudsen number, which justifies the validity of HD/MHD approach. Indeed, as it will be shown in simulations below, the density of planetary protons up to distances of $20 R_p$, exceeds $10^5 cm^{-3}$. Due to the charge-exchange process with a cross-section of $\sim 5 \cdot 10^{-15} cm^2$, the PW protons and hydrogen atoms, while transforming one to another, move practically together with a very insignificant slippage, because their mean-free path is $< 0.2 R_p$. Moreover, the planetary protons are strongly coupled to each other by Coulomb collisions with a cross-section as high as $5 \cdot 10^{-13} cm^2$, calculated at maximum temperature $\sim 10^4$ K of the PW. Due to Coulomb collisions the ions of other heavier species present in the planetary atmosphere are efficiently dragged with PW. Even heavier neutral atoms are coupled sufficiently to be carried along with PW due to elastic collisions with protons with cross-section of order of $10^{-15} cm^2$ which gives the mean-free path $\sim R_p$. Such a strong collisional coupling of species in the expanding PW is important for further production of ENAs.

Then, the expanding PW collides with SW, which is much hotter and faster, though orders of magnitude more rarified and contrary to PW, essentially non-collisional. The coupling of planetary and stellar protons via Coulomb collisions is insignificant because due to the high velocity of SW protons, the mean-free path is of order of $10^{11}$ cm, or $10 R_p$. At the same time, the coupling of ionized species through the magnetic fields should be very efficient for the considered system. Even in the presence of a very weak magnetic field, either regular or chaotic, which is indeed the case in the SW (as well as in PW), the gyroradius and interpenetration depth of the colliding proton flows is orders of magnitude smaller than the typical distance to the sub-stellar pressure balance point of the SW and PW flows. Therefore, a sharp ionopause and a bowshock (in the case of a supersonic SW) have to be formed in the region where the expanding PW meets the SW. If the PW is still a subsonic one, i.e., not accelerated enough before the ionopause due to decrease of pressure difference (a so-called subsonic breeze in *Garcıa Munoz 2007*), it will be simply deflected and redirected towards the nightside and form a kind of a tail.

While the protons of the expanding PW are stopped at ionopause, the planetary atoms penetrate freely through it and enter into the SW. That view differs from the consideration of *Tremblin & Chiang* (*2013*) and *Christie et al.* (*2016*), who supposed the planetary particles to form a single cold fluid. In our case the slippage between the ionized and neutral components of the PW and their respective velocities depend on the plasma conditions within the whole area bounded by the ionopause. After crossing the ionopause, the atoms of planetary origin, because of low collisional rate with the SW particles, move by ballistic trajectories in the SW up to several planetary radii before undergoing the charge-exchange with fast stellar protons and producing the ENAs. The ENAs in their turn can freely pass through the ionopause into the planetary plasmasphere before being ionized either by charge-exchange or stopped by elastic collision with dense PW. In this way an ENA cloud is formed around the planet, mostly in the shocked region between the ionopause and bowshock, because the SW proton density is 4 times larger there. It behaves differently from the stellar protons which simply sweep by along the ionopause in the tailward direction. The sub-stellar part of the ENA cloud expands, due to the thermal pressure, upstream, while at the flanks the ENAs move alongside with the SW. When the ionopause is positioned beyond the first Lagrange point, the pull of stellar gravity becomes crucial. Since the SW pressure scales with orbital distance as $\sim D^{-1}$ (for $D < 0.1$) while the stellar gravitational attraction scales as $\sim D^{-2}$, the PW will be captured by stellar gravity sufficiently close to the star, forming a kind of an accretion flow. The crucial question in that respect concerns the orbital distance at which the SW pressure will become high enough to exceed the tidal force and to push all the PW towards the tail.

We call the above described two principally different scenarios of PW and SW interaction as the *"blown by the wind"* and *"captured by the star"* regimes. The question on which particular regime is realized under given parameters of the stellar-planetary system can be approximately answered, even without detailed numerical simulation of the SW and PW interaction, just by the comparison of the SW total pressure with that of PW, calculated in the case of free expansion of the planetary material. Unimpeded expanding planetary material flow forms a supersonic wind, which beyond the Roche lobe is accelerated further by the tidal potential (7). Because of gravitational collimation of the flow in transverse direction (*Trammel et al. 2011, Khodachenko et al. 2015*) which results in a relatively slow density decrease in the direction towards the star, the total pressure of the modeled PW reaches a minimum value at distances beyond the first Lagrange point (at $\sim 1.5 \cdot R_{L1}$) and then increases slowly due to acceleration by the stellar gravitation. These features are demonstrated in *Figure 2*. If the SW at the corresponding orbital distance has the total pressure below this minimum value, then it couldn't stop the planetary flow, which would be pulled by tidal force, and the *"captured by the star"* regime of the PW and SW interaction will be realized.

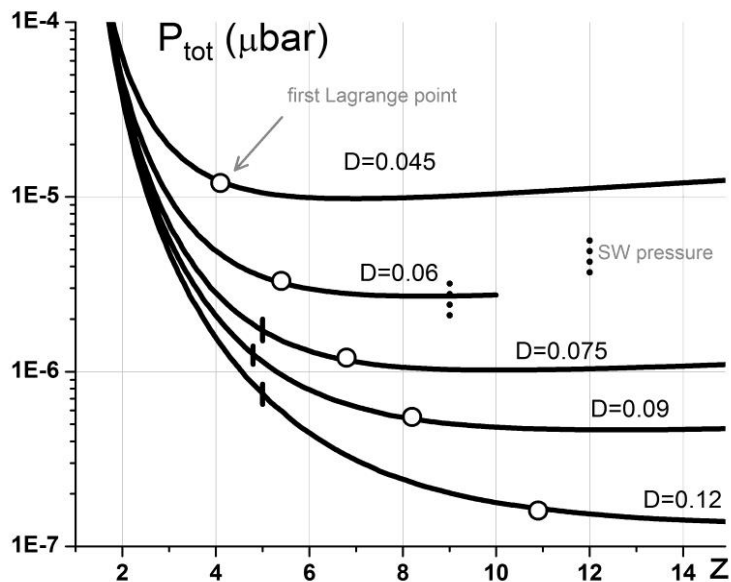

**Figure 2.** Dependence of PW total pressure (sum of ram and thermal pressures) on the height along the planet-star line (in units of $R_p$) obtained for various orbital distances D[AU] in simulations without SW. Circles indicate the corresponding position of the first Lagrange point $R_{L1}$. Vertical bars indicate the range of fast and slow SW total pressure according to parameters in *Table 2*. When possible, the bars are positioned at the corresponding height where the SW and PW pressure balance occurs.

Vertical bars in *Figure 2* indicate the total pressure range between slow and fast SW for each considered orbital distance calculated with data from to the *Table 2*. One can see that at D=0.045 AU, which is the present location of HD209458b, the SW exerts the pressure by a factor of two smaller than the minimum pressure of the unimpeded PW. At D=0.06 AU the SW pressure is barely enough to balance PW, and for larger orbital distances the modeled PW can be stopped by the corresponding SW at distances of about $(5 \pm 0.5) \cdot R_p$ well inside of the Roche lobe. Note that this is still a rough not self-consistent estimation, and the detailed numerical simulation, presented in the following up sections 4.1 and 4.2 reveals insignificant deviation in the parameter range for which particular interaction regime (i.e. *"blown by the wind"* or *"captured by the star"*) is realized. In a similar qualitative study by *Murray-Clay et al.* (*2009*) the SW pressure was taken a bit higher, $10^{-5}$ μbar, which resulted in a balance between the SW and PW of HD209458b around the sonic point height, slightly below $4R_p$.

The qualitative picture of the PW and SW interaction outlined above lacks one important element - the radiation pressure of the star which acts on atomic hydrogen in the Lyα spectral band and is known to affect the dynamics of neutrals in the stellar wind (*Bourrier & Lecavelier des Etangs 2013, Bourrier et al. 2015*). This is certainly a restriction of the model considered here. In support of this limitation, we would like to emphasize, that the subject of the present modelling consists in the simulation of the expanding PW and different regimes of its interaction with the SW sufficiently close to the planet, rather than in the description of dynamics of the lost planetary material in the astrospheric scales. As mentioned in Section 1 (Introduction), the latter needs the engagement of 3D particle codes that include the whole range of relevant physical processes (e.g., charge-exchange, photoionization, and radiation pressure), and extend the hydrodynamic simulation results of present paper to the collision-less regions further away from the planet. We would like to point out that neglecting radiation pressure can be fully justified for the modeling of escaping PW and its consequent interaction with SW. This is supported by the following arguments. First, the considered planetary material flow remains optically thick for Lyα at the line center where $\sigma_{abs} \approx 6 \cdot 10^{-14} \text{cm}^{-2}$. Even at lowest densities $n_H^{pw} \sim 2 \cdot 10^3 \text{cm}^{-3}$ achieved in the simulations far from the planet, the absorption depth is $<1R_p$ and significantly smaller than the size of the whole expanding PW envelope. In this optically thick case the radiation force acts as a surface pressure, which is effectively redistributed, due to collisional coupling, to the PW protons as well. Second, the reconstructed for HD209458b (Vidal-Madjar et al. 2003) Lyα flux in the line center $F_{Ly\alpha} \approx (0.5 - 2) \cdot 10^{14} \text{ph cm}^{-2}\text{s}^{-1}$ at orbital distance of 0.045 AU gives the radiation pressure $P_{Ly\alpha} = F \cdot h\nu/c \approx (0.25 - 1) \cdot 10^{-7} \mu\text{bar}$, which is about two orders of magnitude smaller than the SW ram pressure of a Sun-like star (see in Figure 2). Thus, SW ram pressure significantly exceeds the Lyα radiation pressure in all considered orbital distances up to the stellar corona. This is also the case for the PW stream realized in the *"captured by the star"* regime. Third, besides of the SW pressure, the stellar gravity pull beyond the Roche lobe becomes a dominating force, since the radiation force decreases due to photo-ionization of atoms in the expanding PW. Note, that sufficiently far from the planet the escaping PW becomes strongly photo-ionized.

On the other hand, the rather tenuous flow of planetary atoms decoupled from protons at the ionopause and penetrating into the SW while taking part in generation of the ENA cloud, is optically thin for the Lyα. Therefore this population of particles is affected by the radiation pressure. At a typical orbital position of HD209458b the hydrogen atoms with speeds below 100 km/s absorb Lyα photons with the rate of about $F_{Ly\alpha}\sigma_{abs} \approx 3 \text{ s}^{-1}$ and are ionized by XUV in course of 4-6 h. During this time the radiative pressure may accelerate them up to velocities of 150 km/s (*Vidal-Madjar et al. 2003, Kislyakova et al. 2014a, Bourrier and Lecavelier des Etangs 2013*). Note that beyond 100 km/s the atom leaves the line half-width, and acceleration by radiative pressure practically stops. However, as it will be shown further by simulations, the planetary atoms once decoupled from planetary protons at the ionopause are quickly accelerated to high velocities by pressure gradient and cross the shocked region of $\approx 1R_p$ width in the sub-stellar direction during approximately 1 h. During this time the contribution of the radiation pressure effect to the particle velocity would be of the same order as the initial velocity, therefore it should not change significantly the whole picture of particle dynamics. For the atoms moving in the transverse, relative to the planet-star line, direction across the shocked region, which is of approximately $6R_p$ width, the crossing time is about of 3 h. During this time, while crossing the shocked region, the atoms would be swept by the radiation towards the tail by a distance of about $1R_p$, which is also not significant for the considered global distributions at the scales of several $R_p$. As to the ENAs which are produced mostly in the shocked region where the SW is four times denser, they have thermal or direct velocities larger than 100 km/s and therefore, significantly less absorption rate of Ly$\alpha$, so that the effect of radiation pressure becomes noticeable only at distances of tens of $R_p$.

Finally, the above given arguments justify the neglecting of the radiative pressure in our present study of the PW and SW interaction and related generation of ENAs in the area encompassing ~$10R_p$ around an HD209458b analogue planet. However, the physically more accurate modeling requires the inclusion of the radiation pressure for better quantitative reproduction of the potentially observable features.

## 4. Simulations results

In this section we present the results of modelling of the interaction between the escaping atmosphere of a Hot Jupiter and the incoming SW. As already mentioned in the previous sections, for the definiteness sake we take as a prototype of a Hot Jupiter an analogue of HD 209458b, i.e. a hypothetic exoplanet with the same mass and radius as those of HD209458b, and place it at different distances around a Sun-like star with the solar-type XUV radiation spectrum. The applied numerical code is based on the solver used in our previous modelling of a Hot Jupiter's PW (*Shaikhislamov et al. 2014, Khodachenko et al. 2015*). The explicit second order numerical scheme is employed to treat the model equations, discretized on a cylinder (see in *Figure 1*) uniform mesh with a step of $0.01R_p$ within a simulation domain $(r = 20 \times Z = 40)R_p$. Comparison with our previous much more refined 1D simulations (*Shaikhislamov et al. 2014*) shows that it is sufficient to resolve XUV energy absorption regions in the dense highly stratified atmosphere close to the conventional surface of a planet. As an initial state of the simulated atmosphere we take the pure molecular hydrogen gas with the temperature of 1000 K, distributed according to the static barometric equilibrium. The particle density at the conventional planet surface (i.e. at the inner boundary of the simulation domain $R=R_p$) is fixed at $10^{16}$ cm$^{-3}$, which corresponds to the pressure of about 1000 μbar. Such a large density has been chosen to ensure complete absorption of the XUV flux, so that the particular height of the inner boundary location does not affect the solution (*Shaikhislamov et al. 2014, Khodachenko et al. 2015*). An SW, approaching the planet from the stellar side is applied as an outer boundary condition.

The major physical parameters and results of the simulations further on will be scaled in units of the characteristic values of the problem, defined as follows: for temperature, $T_0 = 10^4$ K; for speed, $V_0 = V_{Ti}(T_0) = 9.1$ km/s; for distance, the planet radius $R_p$; for time, the typical flow time $t_0 = R_p/V_0 \approx 3$ hr. A stationary solution is sought with convergence criteria defined as the relative change of calculated values (e.g., the mass loss) at the outer boundary at one step not exceeding $10^{-2}$. Usually it takes several hundred dimensionless times or $\sim 10^6$ iterations after the onset of the planet heating by XUV to achieve the steady state.

### 4.1 *The "blown by the wind" regime of the PW and SW interaction*

At first we consider the case when the SW total pressure is sufficiently high to stop the PW and to form an ionopause and bowshock. Within the above used terminology, this case corresponds to the *"blown by the wind"* regime of the PW and SW interaction. For the considered analogue of HD209458b, this regime is realized at an orbital distance of D=0.075 AU under the conditions of fast SW. *Figure 3* shows the density distributions of the main interacting fluids – protons and hydrogen atoms of planetary and stellar origin, respectively. The stellar hydrogen atoms, produced due to charge-exchange between the SW protons and neutrals of the PW, form the ENA cloud (the bottom-right panel in *Figure 3*). The motion of each fluid is marked by corresponding streamlines. In the distributions of $n_{H+}^{pw}$ and $n_{H+}^{sw}$ in the upper panels one can see a sharp bowshock, formed due to coupling of PW and SW protons. In the shocked region between the ionopause and the bowshock, the stellar protons are deflected and flow around the planet (see the top-right panel in *Figure 3*). The whole PW region is bounded by the ionopause, at which the planetary ions collide with stellar protons. By this, the planetary ions are deflected around the planet and redirected towards the tail. One can see also a turbulent character of the sheared flow region near the ionopause which produces vortices due to interchange instability (e.g. Kelvin-Helmholtz). The distribution of planetary atoms is essentially different from that of planetary ions. Their density sharply decreases at ionopause, like that of ions, but they penetrate freely far into the stellar wind. Below the ionopause planetary atoms are strongly coupled to planetary ions by charge-exchange while in the shocked region and further on in the stellar wind the density of particles is too low for significant coupling. This is reflected in the specific shape of streamlines of the planetary atoms at the day side (see the bottom-left panel in *Figure 3*) which are dragged by ions around the ionopause and after that move practically along the straight ballistic trajectories. As it can be seen in the plot, the injection of planetary atoms across the ionopause is spatially modulated by streaming blobs formed by instabilities. The bottom-right panel in *Figure 3* shows ENA cloud which has a typical density of the order of a few particles per cm$^{-3}$. It is located mainly between the bowshock and the ionopause (because in this region a product of densities of planetary atoms and stellar protons has a sharp maximum).

Note, that the ENA streamlines are close to those of the stellar protons, but not identical. The ENA cloud, localized mainly in the region between the ionopause and bowshock, expands, due to the particles thermal motion, into the upstream stellar wind. In the bottom-right panel in *Figure 3* one can also see a stagnant population of neutral hydrogen atoms of stellar origin accumulated around the planet, mainly on the nightside. Their temperature and velocity are equal to those of the planetary wind, and in spite of being originated at the star, these particles are similar to the majority of the local cold population of the planetary origin. In fact, they have been created at the beginning of the simulation process, during the initial stage of the modelled planetary plasmasphere formation, when the escaping PW was still under formation and certain amount of the SW protons was able to penetrate deep into the planetary atmosphere before the quasy-stationary picture of the PW and SW interaction with the ionopause and bowshock has been established. These cold neutral hydrogen atoms of stellar origin can be easy neglected, since they constitute an extremely small fraction in the majority of the planetary atmospheric hydrogen in the inner PW.

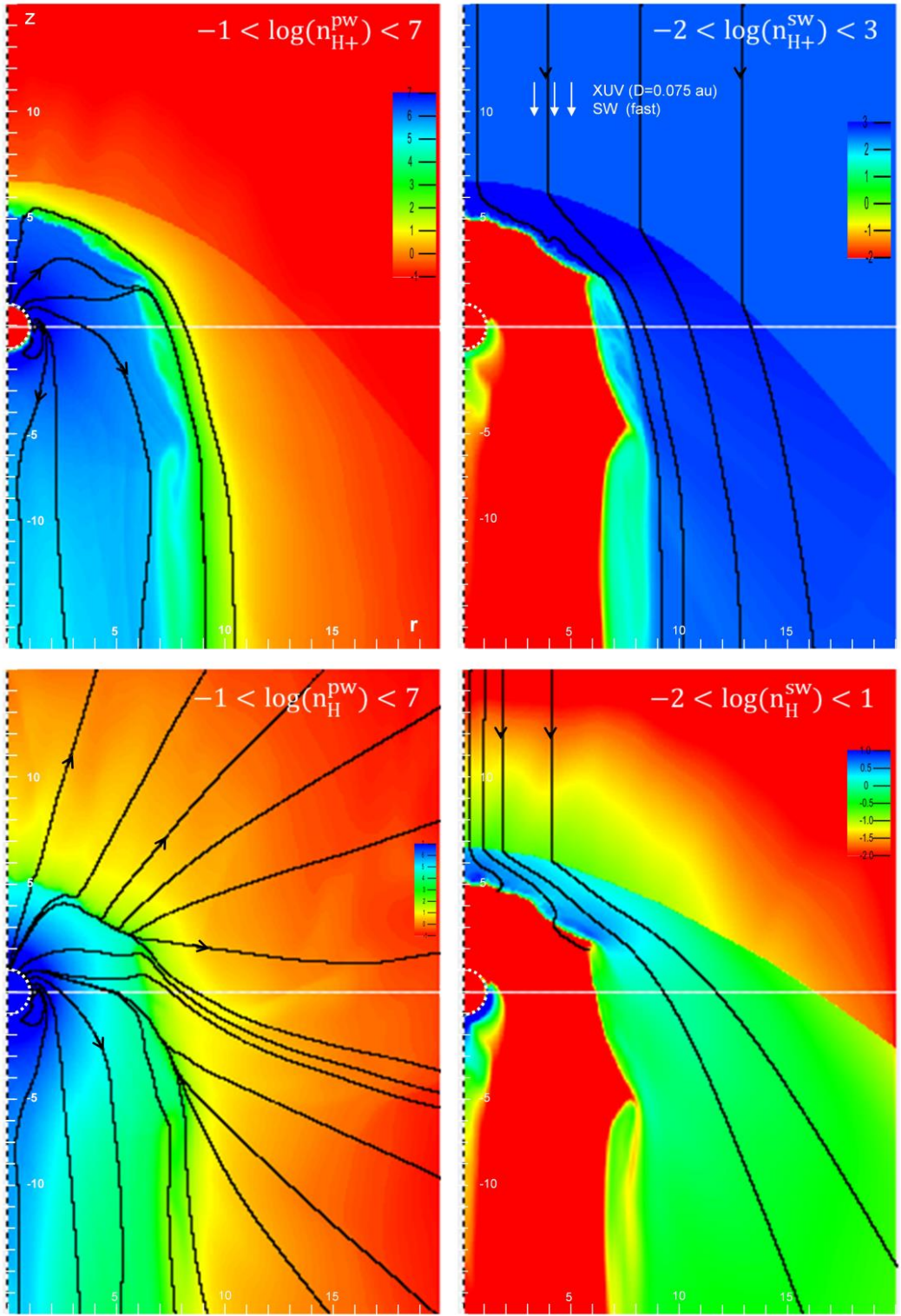

**Figure 3.** Density distributions of planetary protons ($n_{H+}^{pw}$), stellar protons ($n_{H+}^{sw}$), planetary atoms ($n_{H}^{pw}$), and ENAs ($n_{H}^{sw}$), realized during the interaction of fast SW with the expanding PW of an analog of HD209458b at orbital distance of 0.075 AU (*"blown by the wind" regime*). The plotted values are in log scale. The streamlines of the corresponding components are shown in black. Here and further on, the values outside the indicated variation ranges of the plotted parameters are colored either in red if smaller than minimum, or in blue, if higher than maximum. White dashed line circle indicates the planet.

The first two panels in *Figure 4* show the temperature distributions (in units of $10^4$K) of the interacting PW and SW components. The electron temperature is equal to that of ions, and the left panel in *Figure 4* reflects the hot stellar wind temperature range, while the middle panel shows the temperature of much colder planetary atomic hydrogen inside the ionopause. The right panel in *Figure 4* shows the content of atomic hydrogen in the planetary material flow. One can see that already at the ionopause it is rather small ~4%. In the shocked region it is greatly increased in the vortex-type streams formed by instabilities. The color scale has been chosen so that the details of the PW atomic hydrogen distribution in the shocked region are clearly visible.

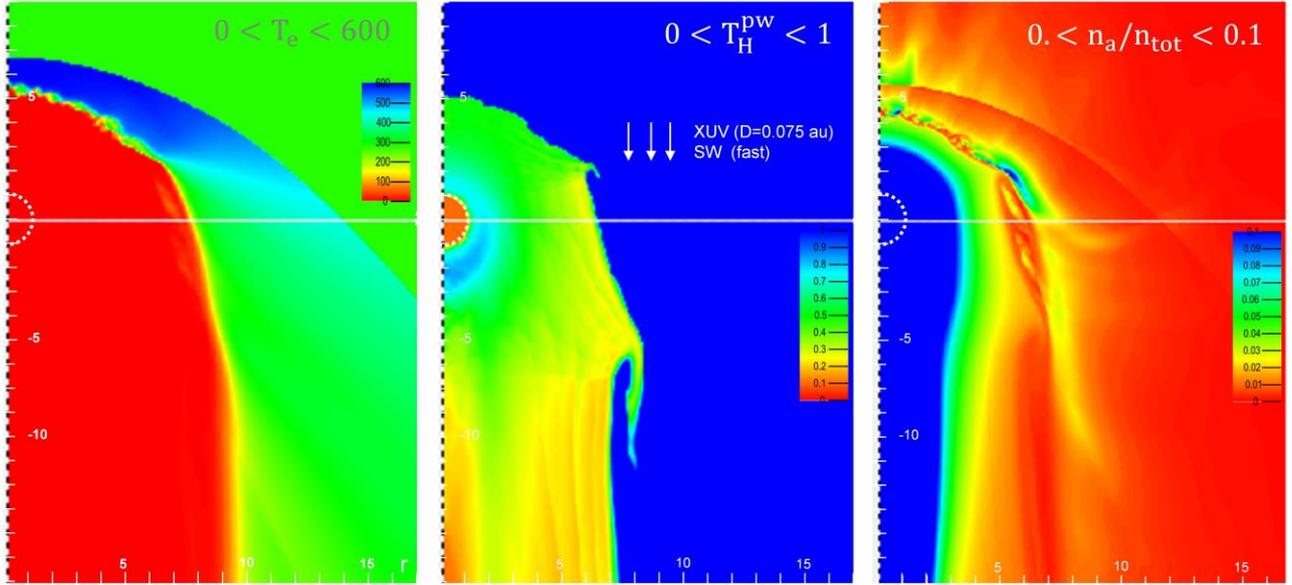

**Figure 4.** Spatial distribution (from left to right) of electron temperature $T_e$, temperature of planetary hydrogen atoms $T_H^{pw}$, and the number densities ratio of planetary atoms to that of all hydrogen particles in PW realized during the interaction of fast SW with the expanding PW of an analog of HD209458b at orbital distance of 0.075 AU (*"blown by the wind" regime*).

The pressure profiles in the sub-stellar direction, shown in *Figure 5*, demonstrate the structure of the shock. At the bowshock, located at about $6.8R_p$, the SW thermal pressure increases sharply and becomes comparable to the upsream ram pressure of SW. The corresponding jump of density by 4.4 times (see in *Figure 6*) is close to the Rankine–Hugoniot relation. At the ionopause positioned at about $5.5R_p$ the PW total pressure, contributed mostly by the thermal pressure of ions, balances the SW total pressure.

The profiles of density, velocity and temperature of the interacting PW and SW components, shown in *Figure 6* reveal that close to ionopause from the planetary side the PW is strongly ionized with ion density prevailing by more than order of magnitude, because of the ionization of neutrals by the stellar XUV. However, in the shocked region and in the stellar wind the planetary atoms dominate by more than an order of magnitude. The planetary ions exist in these regions only as a by-product of charge-exchange of planetary atoms with stellar protons. Velocity profiles (middle panel in *Figure 6*) demonstrate strong coupling of planetary and stellar protons assumed in the model and strong coupling of planetary ions and atoms in the inner PW region (below the ionopause) existing due to charge-exchange. The velocity of PW itself is rather small not exceeding the value of 2 km/s which indeed corresponds to the regime of a sub-sonic breeze expansion. Radial velocity of planetary particles in the region below the ionopause is quite small as compared to the free expansion case, but there is significant zonal flow velocity which transports material towards the nightside and to the tail. The small radial velocity implies a longer time for the planetary atom to travel from the dense thermosphere to the ionopause, resulting in a larger photo-ionization probability.

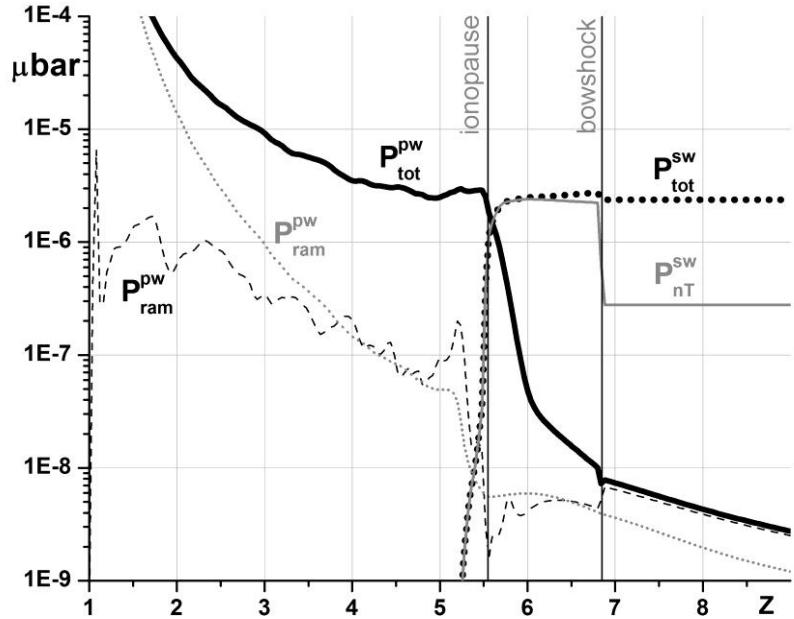

**Figure 5.** Pressure profiles of the interacting PW and SW along the planet-star line for HD209458b analogue at orbital distance D=0.075 AU, and fast SW (*"blown by the wind" regime*): total pressure of PW (black solid), total pressure of SW (black dot), thermal pressure of SW (grey solid), total pressure of planetary atoms (gray dot) and PW ram pressure (black dash).

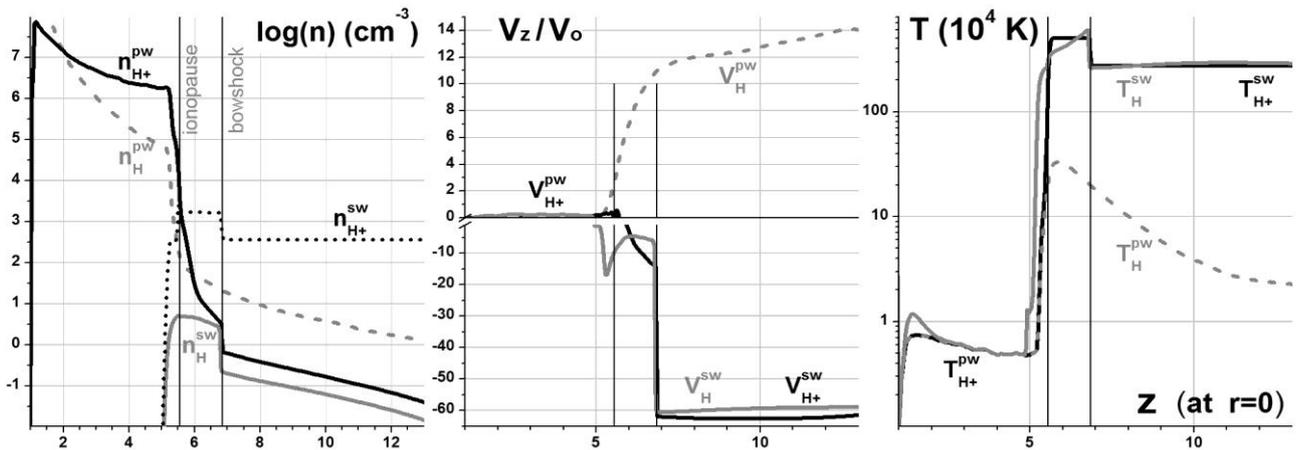

**Figure 6.** Profiles of densities (left panel), velocities (middle) and temperatures (right) for planetary ions (black solid), SW protons (black dot), planetary atoms (gray dash) and ENAs (gray solid) along the planet-star line for HD209458b analogue at orbital distance D=0.075 AU, and fast SW (*"blown by the wind"* regime).

Immediately beyond the ionopause the density of PW ions becomes sufficiently small, so the planetary atoms decouple and are sharply accelerated up to ~100 km/s by the pressure gradient. Such acceleration is supported by the simultaneous temperature increase of the injected planetary atoms due to rare elastic collisions with extremely hot SW protons in the shocked region. Note, that after the shocked region, the adiabatic cooling of the expanding planetary atoms dominates over the heating (right panel in *Figure 6*).

The radial profiles of densities and temperatures across the planet-star line at z=0 in *Figure 7* reveal that the bowshock of supersonic ENA gas appears at a slightly different position than the stellar protons bowshock. This is explained by different structure of ENA cloud and its expansion from the production region. The temperatures of planetary atoms and ions are practically the same in the inner region with a maximum value slightly short of $10^4$ K. However, in the rarified region beyond the ionopause the temperatures decouple and planetary protons are heated to the temperature of stellar protons, while planetary atoms are heated only slightly (due to rare elastic collisions with hot SW protons) and cool with distance due to adiabatic expansion.

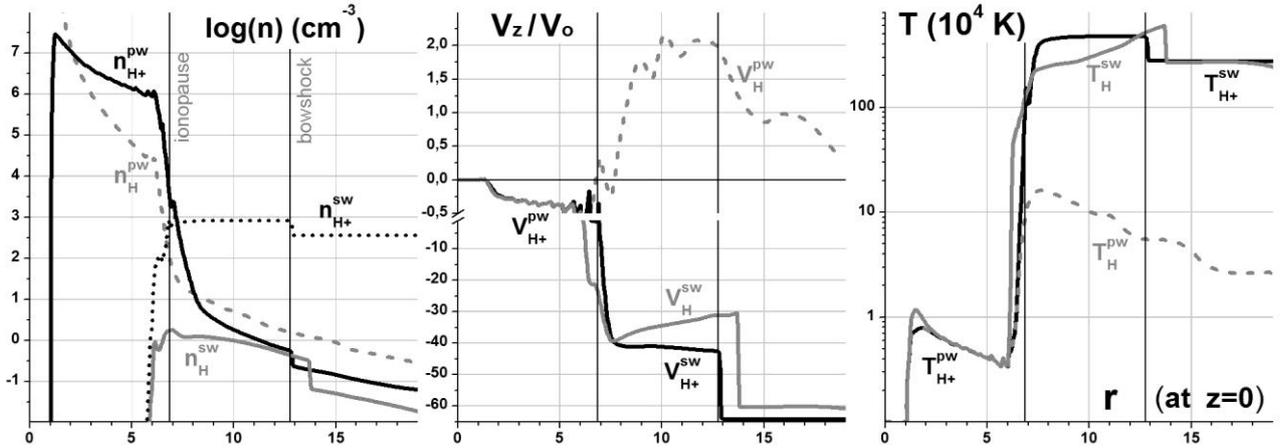

**Figure 7.** Profiles of densities (left panel), velocities (middle) and temperatures (right) for planetary ions (black solid), SW protons (black dot), planetary atoms (gray dash) and ENAs (gray solid) across the planet-star line at z=0 for HD209458b analogue at orbital distance D=0.075 AU, and fast SW (*"blown by the wind" regime*).

Nearly the same picture was observed at larger orbital distances, D=0.09 AU and D=0.12 AU for both fast and slow SW. For example, *Figures 8* and *9* show the density, velocity and temperature distributions of the PW and SW components including the ENAs, realized during the interaction of slow SW with the expanding PW of an analog of HD209458b at orbital distance of 0.12 AU. The positions of the ionopause and bowshock agree with the qualitative prediction in *Figure 2*. Note, that ENAs again are generated in the shocked region. Because of two times larger density of the slow SW and higher charge-exchange cross-section at lower velocities, the density of ENA cloud is several times larger than that in the case of fast SW. The ENA density can be estimated by balancing the convection and production terms in the continuity equation:

$$V_H^{sw} n_H^{sw}/L \sim n_{H+}^{pw} n_{H+}^{sw} \sigma_{exh} V_{H+}^{sw}, \rightarrow n_H^{sw} \sim n_{H+}^{sw} n_H^{pw} L \sigma_{exh} \qquad . \qquad (10)$$

Here L is a typical gradient scale, which is of the order of width of the shocked region. Taking $n_{H+}^{sw} = 10^3 \mathrm{cm}^{-3}$, $L = 5 R_p$, $n_H^{pw} \sim 20 \mathrm{cm}^{-3}$, as can be judged from the *Figure 9*, and $\sigma_{exh} = 2 \cdot 10^{-15} \mathrm{cm}^2$, one can obtain $n_H^{sw} \sim 3 \mathrm{cm}^{-3}$, which is rather close to the value observed in simulations. The largest uncertainty is in a value of the planetary atoms density $n_H^{pw}$ in the shocked region which is difficult to derive analytically. The estimations show, that recombination with the rate of $2.5 \cdot 10^{-13} n_e T_e^{-0.75} [10^4 K] s^{-1}$ at typical electron density of $10^6$ cm$^{-3}$ and $T_e \sim 5000$ K can balance XUV ionization at this orbital distance D=0.12 AU only at a high ionization degree of about 99% which is significantly larger than that observed in our numerical simulation (*e.g., see in Figure 9*). Instead, as it follows from equation (1), the atomic density is given by the photo-ionization rate multiplied by advection time, i.e. $n_H^{pw} \sim \exp\left(-\int \Gamma_{ion} dl / V\right)$ and depends on the velocity in whole inner region.

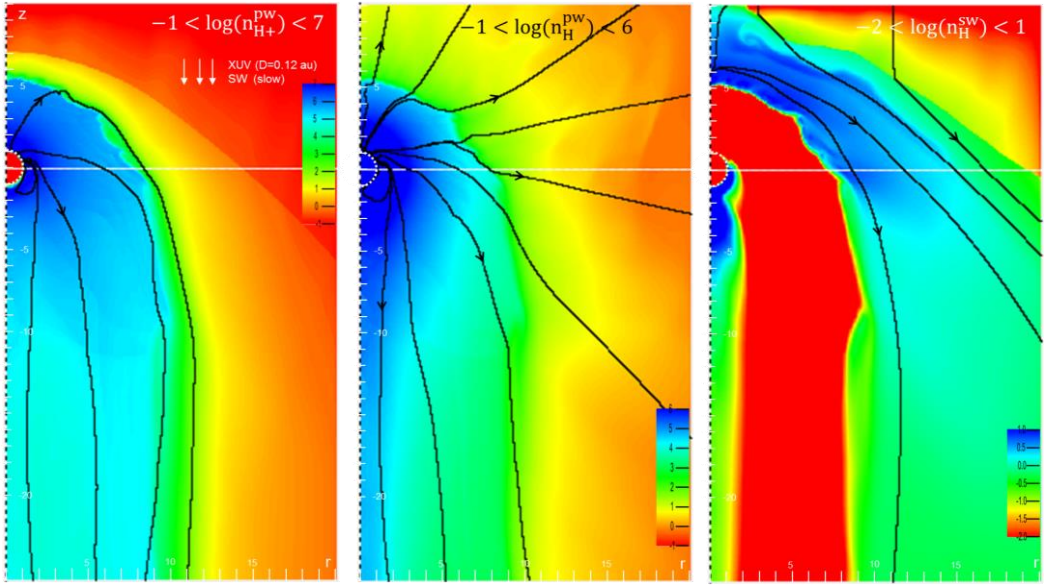

**Figure 8.** Density distributions of planetary protons ($n_{H+}^{pw}$), planetary atoms ($n_H^{pw}$), and ENAs ($n_H^{sw}$), realized during the interaction of slow SW with the expanding PW of an analog of HD209458b at orbital distance of 0.12 AU (*"blown by the wind"* regime). The plotted values are in log scale. The streamlines of the corresponding components are shown in black. White dashed line circle indicates the planet.

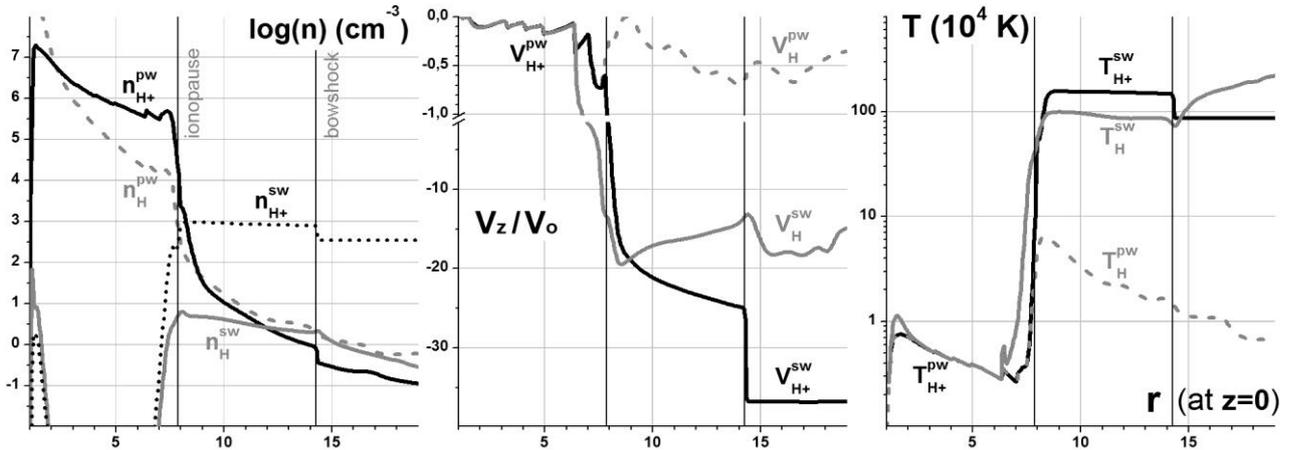

**Figure 9.** Profiles of density (left panel), velocity (middle panel), and temperature (right panel) for planetary ions (black solid), SW protons (black dotted), planetary atoms (gray dashed), and ENAs (gray solid) across the planet-star line at z=0 in the case of HD209458b analogue at orbital distance D=0.12 AU, and slow SW (*"blown by the wind"* regime).

## 4.2 The "captured by the star" regime of the PW and SW interaction

It was found in simulations, that while fast SW stops the PW of an analogue of HD209458b at the orbital distance of D=0.075 AU, the slow SW does not, and the PW expands towards the star, beyond the Roche lobe, where it is pulled further by the tidal force. Within the above used terminology, such a situation corresponds to the *"captured by the star"* regime of the PW and SW interaction. *Figure 10* shows the color plots of the densities of the interacting SW and PW components in this case. One can see that planetary protons expand towards the star much beyond the first Lagrange point ($R_{Lag,1} = 6.8R_p$) and are not stopped in their propagation at all, although the thermal pressure of the SW is strong enough to squeeze the stream of the planetary material from the lateral side. A kind of an ionopause boundary is formed between the stream of the escaping PW and the surrounding SW flow.

Some of the planetary protons, passing close to the lateral boundary of the planetary material stream are picked up by the SW flow and deflected towards the tail (see the top-left panel in *Figure 10*). However, the number of these particles is very small and the downstream flux of the reflected planetary protons is 2 orders of magnitude less than the ionized material flow towards the star. On the other hand, the planetary atoms traveling along the same streamlines are not deflected at the ionopause and penetrate directly into SW flow, as they exchange momentum with the SW protons only weakly (see the bottom-left panel in *Figure 10*). The ENA cloud in the case of a *"captured by the star"* regime shows a principal difference from that formed during the *"blown by the wind"* type of interaction between PW and SW. Indeed, due to the effect of the tidal force (*Trammell et al. 2011, Khodachenko et al. 2015*), majority of planetary atoms (until being photo-ionized) flow inside the stream tube along the planet-star line together with planetary protons and do not have a significant transverse velocity to penetrate across the tube boundary into the SW. Because of that, there is no sufficiently large region where the planetary atoms could meet the SW protons and efficiently interact with them by charge-exchange giving the origin for ENAs. As a result, the density of ENAs around the planet in the case of a *"captured by the star"* regime of the PW and SW interaction realized for an HD209458b analog in a slow SW, is nearly an order of magnitude smaller than that in the case of the *"blown by the wind"* regime under similar planetary conditions (orbital distance) but for the fast SW (even despite of the higher SW density and larger charge-exchange cross section in the slow SW). In fact, in the *"captured by the star"* regime the mixing of the planetary atoms with stellar protons, which provides the conditions for charge-exchange interaction, takes place (similarly to that shown in *Tremblin & Chiang* (*2013*) and *Christie et al.* (*2016*)) via interchange instabilities at the boundary of the PW and SW flows, rather than due to direct penetration of planetary atoms into SW. Like in the large-scale modelling of the whole star-planet system by *Matsakos et al.* (*2015*), in our simulation the boundary of accretion stream exhibits a complex fragmented morphology that arises from the tangential velocity shear.

There is another detail that could be seen in *Figure 10* - the unimpeded planetary flow at the front and/or in the tail quickly becomes a highly supersonic one and might develop the non-stationary shocks even without being actually stopped by SW. In particular, the almost one-dimensional super-sonic flow confined within the stream tube can generate shocks and subsequent turbulence at places of the tube squeezing (see for example in top-left panels in *Figures 10, 11*).

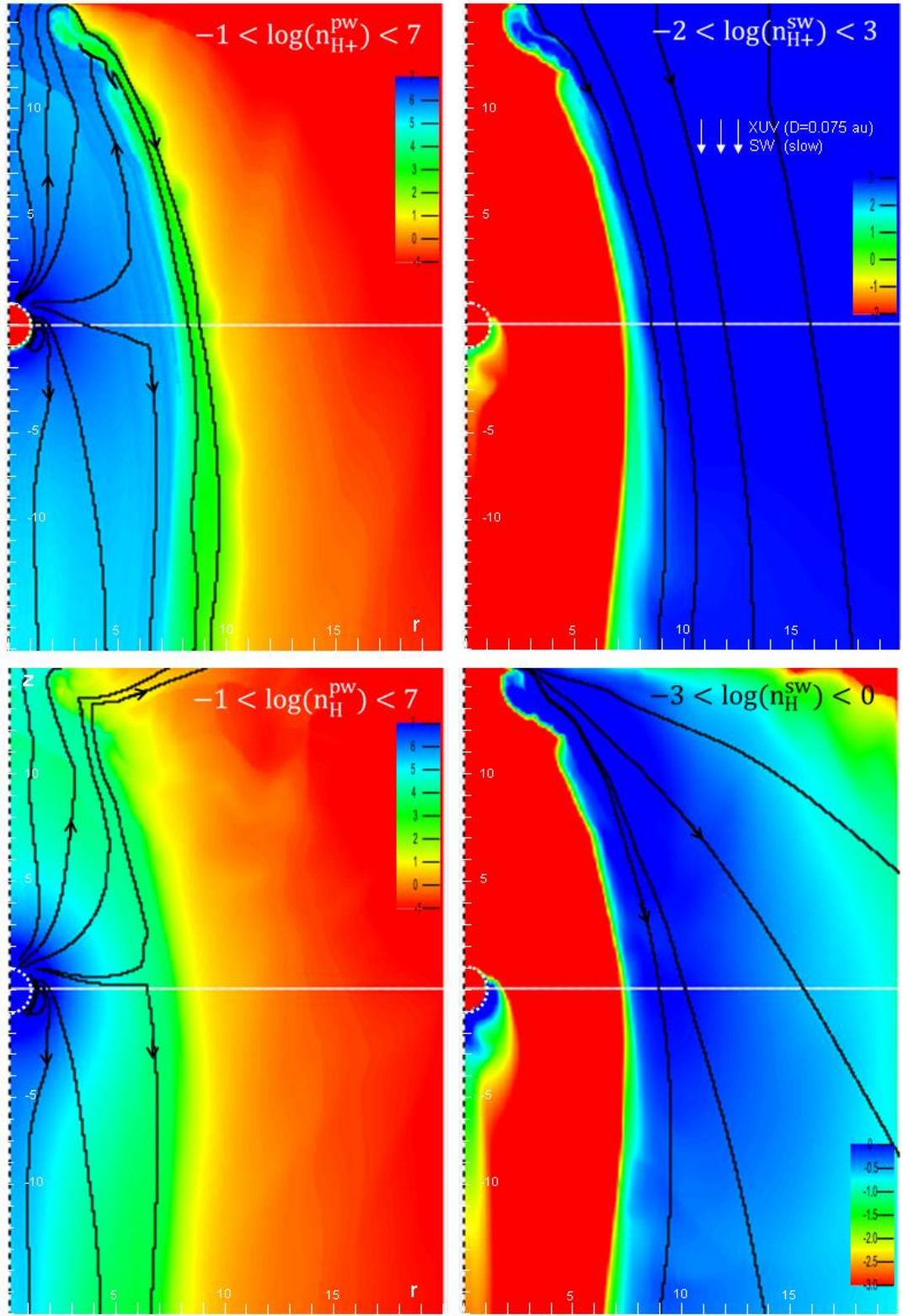

**Figure 10.** Density distributions of planetary protons ($n_{H+}^{pw}$), stellar protons ($n_{H+}^{sw}$), planetary atoms ($n_H^{pw}$), and ENAs ($n_H^{sw}$), realized during the interaction of slow SW with the expanding PW of an analog of HD209458b at orbital distance of 0.075 AU (*"captured by the star"* regime). The plotted values are in log scale. The streamlines of the corresponding components are shown in black. White dashed line circle indicates the planet.

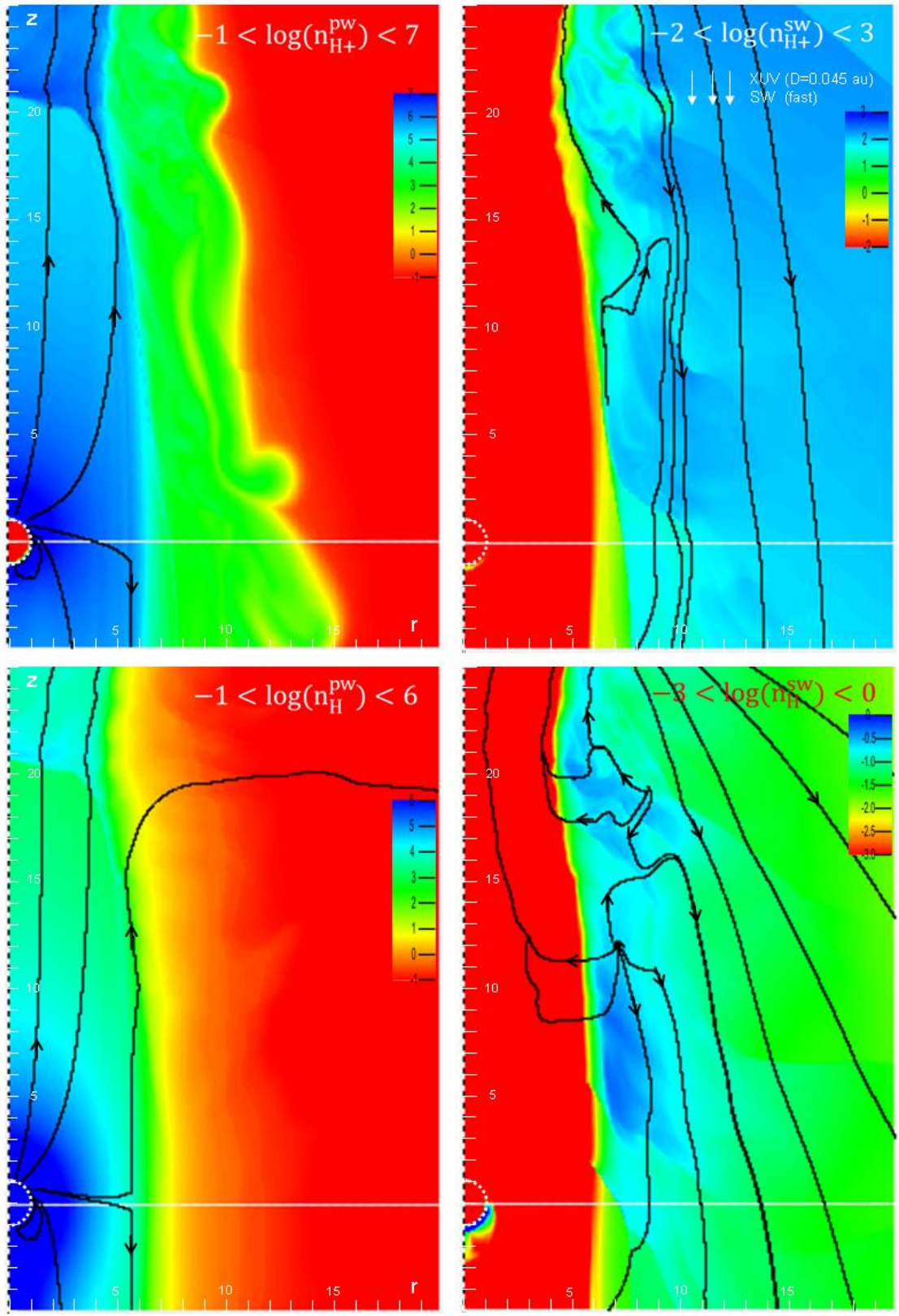

**Figure 11.** Density distributions of planetary protons ($n_{H+}^{pw}$), stellar protons ($n_{H+}^{sw}$), planetary atoms ($n_{H}^{pw}$), and ENAs ($n_{H}^{sw}$), realized during the interaction of fast SW with the expanding PW of an analog of HD209458b at orbital distance of 0.045 AU (*"captured by the star"* regime). The plotted values are in log scale. The streamlines of the corresponding components are shown in black. White dashed line circle indicates the planet.

At closer orbits the gravitational interaction and SW conditions are so, that the escaping PW of an HD209458b analog planet cannot be stopped by the SW, and a fully developed *"captured by the star"* regime of the PW and SW interaction is realized. For example, *Figures 11,12,13* show the distribution of major components of the interacting SW and PW flows as well as their velocity and temperature in the case of an HD209458b analog planet at orbital distance D=0.045 AU and fast SW. One can see that the PW continuously streams along the planet-star line towards the star, reaching the area where its motion is totally driven by the stellar gravity. By this the flow is accelerated up to velocities of almost 100 km/s.

Due to the lateral confinement, the density of the flow decreases slowly, and at a distance of $20R_p$ it still exceeds the SW density by two orders of magnitude. Close to the simulation box boundary a highly supersonic flow ($M_s$>10) terminates in an non-stationary shock with a sub-sonic flow upstream. In the close vicinity outside the planetary material stream the particles density is 3 orders magnitude smaller. In this turbulent transition layer the velocity changes sign and planetary and stellar protons intermix. The majority of planetary atoms, as well as protons, flow mostly inside the stream tube, except of the boundary, where some atoms penetrate into the turbulent transition layer and pass into the SW. However, this amount of planetary atoms is rather small and the resulting ENA density is smaller by an order of magnitude, as compared to the case of *"blown by the wind"* regime. As it was pointed out above, the ENAs are produced at localized centers of mixing, modulated by interchange instability. Since the velocity of stellar protons in the turbulent transition layer is not high, the generated ENAs expand in all directions from these centers of their production, including the direction towards the star (i.e. oppositely to the entire SW).

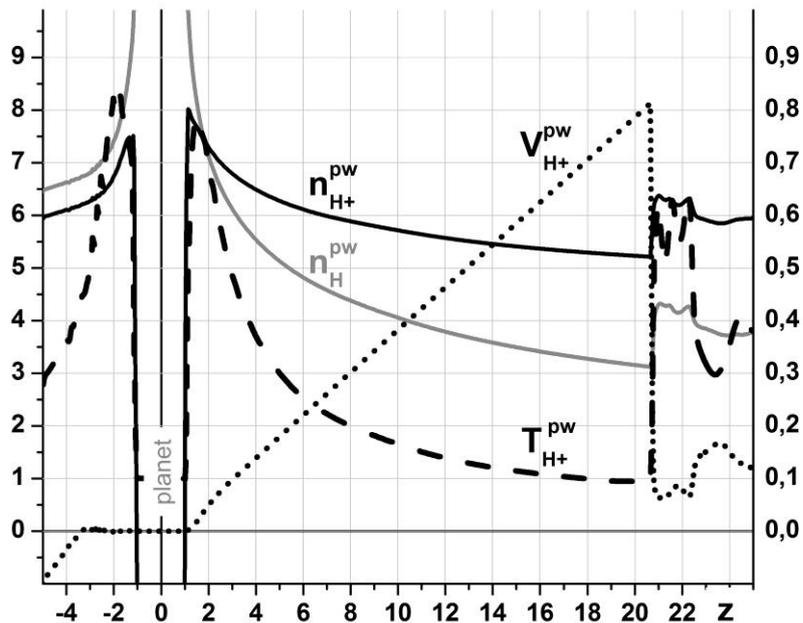

**Figure 12.** PW structure along the planet-star line in the case of HD209458b analogue at orbital distance D=0.045 AU, and fast SW (*"captured by the star" regime*). Left axis: proton density (black solid), hydrogen atom density (gray solid), proton velocity (dotted); Right axis: proton temperature (dashed).

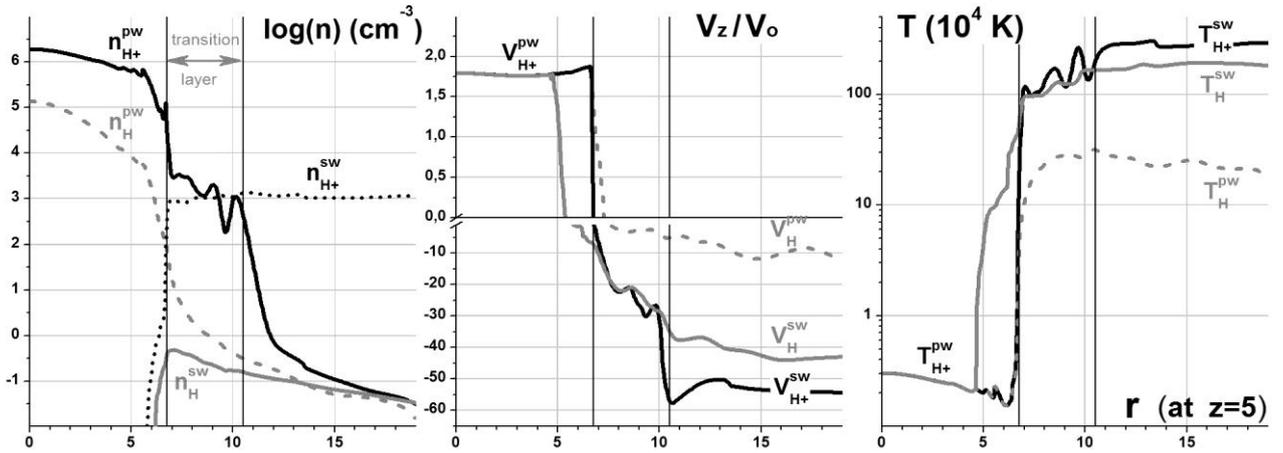

**Figure 13.** Profiles of density (left panel), velocity (middle panel), and temperature (right panel) for planetary ions (black solid), SW protons (black dotted), planetary atoms (gray dashed), and ENAs (gray solid) across the planet-star line at z=5 in the case of HD209458b analogue at orbital distance D=0.045 AU, and fast SW (*"captured by the star"* regime).

### 4.3 *The influence of SW on the Hot Jupiter's mass loss rate*

The mass loss rate is one of the important characteristics of an exoplanetary system, traditionally addressed in the context of the stellar-planetary interaction, planetary evolution, and upper atmospheric material escape. In previous works, we used our numerical model to investigate the dependence of the mass loss rate of an HD209458b analogue planet on the stellar XUV flux (*Shaikhislamov et al. 2014*) and magnetic field (*Khodachenko et al. 2015*). It has been defined as a flux of planetary atoms and ions across the boundaries of the simulation domain. The improved modelling approach presented in this paper allows the quantitative characterization of the effect of a SW on the integral mass loss rate of a Hot Jupiter under different PW and SW interaction regimes. The outcome of this study for an analog of HD209458b planet is summarized in *Figure 14*. The simulated mass loss rate of the considered Hot Jupiter at D=0.045 AU without inclusion of SW is of about $6 \cdot 10^{10}$ g/s. Because of the account of infrared cooling, it decreases with orbital distance a bit more slowly than the XUV flux does ($\sim 1/D^2$).

The net effect caused by the SW is, as expected, quite small. The largest difference, not exceeding a factor of two, is observed at the highest considered distance of D=0.12 AU for the fast SW. This effect is mainly related with the fact that at larger orbits e.g. D > 0.075 AU, the ionopause is essentially below the Roche lobe distance and the *"blown by the wind"* regime is realized for both kinds of SW. In this case, less amount of PW particles, which have to overcome not only the gravity of planet, but also the resistance of the SW flow, are able to escape the planet and become lost. For smaller orbital distances, the difference in the net mass loss rate caused by SW is insignificant because the sufficiently strong PW is realized for both *"captured by the star"* and *"blown by the wind"* regimes of planetary material outflow, and most of the planetary material appears in the escaping regime. This is also connected with the fact, that the nightside flow, which is practically unaffected by the SW, contributes about a half of the integral mass loss. As it was shown in simulations by *Khodachenko et al.* (*2015*), the nature of the nightside PW is connected with a zonal flow which transports the material heated on dayside, to the dusk.

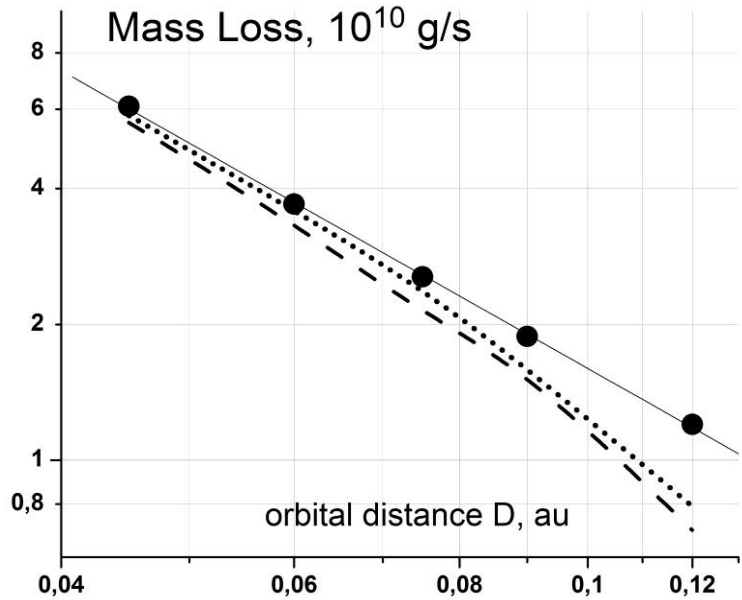

**Figure 14.** The integral mass loss rate of an HD209458b analog planet versus orbital distance without SW (solid), slow SW (dotted) and fast SW (dashed).

## 5. Discussion and conclusions

The presented self-consistent multi-fluid HD simulations of the interaction of the expanding PW of a Hot Jupiter with the upcoming SW confirm the possibility of two principally different regimes of the winds interaction: 1) the *"captured by the star"* and 2) the *"blown by the wind"*. So far, such regimes were described only generally (*Matsakos et al. 2015*), whereas our modelling enables for the first time the detailed description of the structuring of the interaction region and quantifies the related process of the ENA cloud generation and distribution around the planet.

It has been found for a non-magnetized HD209458b analogue planet, that beyond the orbital distance D=0.075 AU, the *"blown by the wind"* regime of the PW and SW interaction is realized, and a kind of pressure dominated plasmosphere is formed which is bounded by ionopause with a sub-stellar pressure balance point located at about $5R_p$. Contrary to previous works we found that because of the SW barrier, the dayside PW does not reach a supersonic regime, and its velocity remains rather low. The flow is redirected completely towards the tail, and the integral mass loss remains just a little bit less than that without the SW. In line with our theoretical expectations, we found that strongly coupled collisionally, the neutral and ionized components of the escaping PW, decouple at the ionopause, where planetary and stellar protons stop each other, whereas the neutral atoms are strongly accelerated by pressure gradient and penetrate into the shocked region and the SW. They interact with the SW protons there by charge-exchange and give the origin for the ENA cloud. The ENA cloud with density of the order of 1 particle per $cm^{-3}$ is produced by charge-exchange mostly in the shocked region (i.e. between the ionopause and bowshock) where the SW proton density increases by several times. At the substellar point of the bowshock the proton velocity falls down to ¼ of its up-stream value and to gradually higher values along the shock extension in the lateral direction up to roughly 5Rp. The predicted by our modelling ENAs have the observer directed velocities in the range 60-100 km/s and 130-200 km/s for slow and fast SW, respectively. It is close to the range of velocities up to -130 km/s where the blue shifted absorption has been observed (*Vidal-Madjar et al. 2008*). This absorption has to have also very significant Doppler broadening, as the thermal velocity in the shock is of the same order of magnitude. It is worth to mention that this feature of the ENA cloud distribution was overlooked so far in the Monte-Carlo simulations (*Holmstrom et al. 2008, Kislyakova et al. 2013, 2014a, Lammer et al. 2011, 2013*), which did not take the presence of the shocked region into account.

At orbital distances below D=0.075 AU the parameters of the HD209458b analogue system result in the *"captured by the star"* regime of the PW and SW interaction, when the SW cannot balance the stellar gravitational pull on the escaping planetary plasma. In this case, the rate of ENAs production appears about an order of magnitude less, as compared to the case of the *"blown by the wind"* regime. The planetary atoms penetrate into the SW, where they can take part in the production of ENA by charge exchange, only due to a kind of interchange instability developing at the boundary of sliding relative each other PW and SW flows. This is analogous but not similar to the mechanism modeled in *Tremblin & Chiang (2013)*. Since the effects of large-scale spiraling of the escaped planetary material due to Coriolis force are not included in our model, the obtained results regarding the distribution of plasma and ENAs could be assumed correct only sufficiently close to the planet, i.e. within the distance range comparable to the stream width $< 10R_p$ in either direction along the planet-star line. At the same time, our simulations show that the PW flow, that overcomes the SW pressure, is accelerated at such distances up to velocities of 30 km/s. This is too small value to be observable in Ly$\alpha$ because of interstellar absorption; however, it might be visible in lines of heavier species. As the PW remains a collisional flow, the heavier ions and atoms will be dragged along with the whole planetary material, and this can explain data on the observed red-shifted absorption of CII, OI mentioned above (*Linsky et al. 2010, Ben-Jaffel and Sona Hosseini 2010*).

The detailed self-consistent simulation of expansion of a Hot Jupiter's PW and injection of planetary atoms into the SW flow with consequent generation of the ENA cloud is the main novelty of the present work. The obtained results show that the scenarios of *Tremblin & Chiang (2013) and Christie et al. (2016)* in which planetary protons and atoms move and mix with SW together are unrealistic, whereas the account of decoupling of the PW atoms from the protons at the ionopause changes the scales and shape of the ENA cloud formed around the planet. Our approach required taking into account of complete multi-fluid physics of the planetary upper atmosphere escape. The performed modeling shows that the amount of atoms that survive photo-ionization until reaching the ionopause is in fact quite small. Note from *Figures 6, 7, 12, 13* that for all PW and SW interaction regimes a half-ionization point is reached slightly below $2R_p$ while at $3R_p$ the number of atoms is about 15%, and at the ionopause (see *Figures 4, 6, 7*) it is only 4%. When the SW forms a barrier (the case of *"blown by the wind"* regime), the radial velocity of PW atoms at the dayside front decreases while the time of exposure to XUV increases, and the ionization degree becomes higher than that in the case of a relatively free expansion under the *"captured by the star"* regime at the same distances below the ionopause (compare *Figures 6 and 12*). By this, the PW flow remains sub-sonic in the *"blown by the wind"* regime of interaction, and this is also a novel feature found in the present work.

Comparison of different works dedicated to the study of PW generation (e.g., *Koskinen et al. 2013, Garcia Munoz 2007, Murray Clay et al. 2009*) shows that there is significant diversity in the radius of half-ionization, predicted for HD209458b by numerical models, varying from $1.6R_p$ to $3.5R_p$. This comes from either specific stellar XUV spectrum models, or different atmospheric composition and particular boundary conditions. It has crucial consequences for the modelling of the ENA cloud and prediction of its absorption properties at high velocity wings of Ly$\alpha$ line, which are important for interpretation of the observed HD209458b transit spectra. For example, a smaller density of planetary neutrals which penetrate through the ionopause into SW results in a less dense ENA cloud which in its turn cannot provide any significant (detectable) Ly$\alpha$ absorption during the planetary transit.

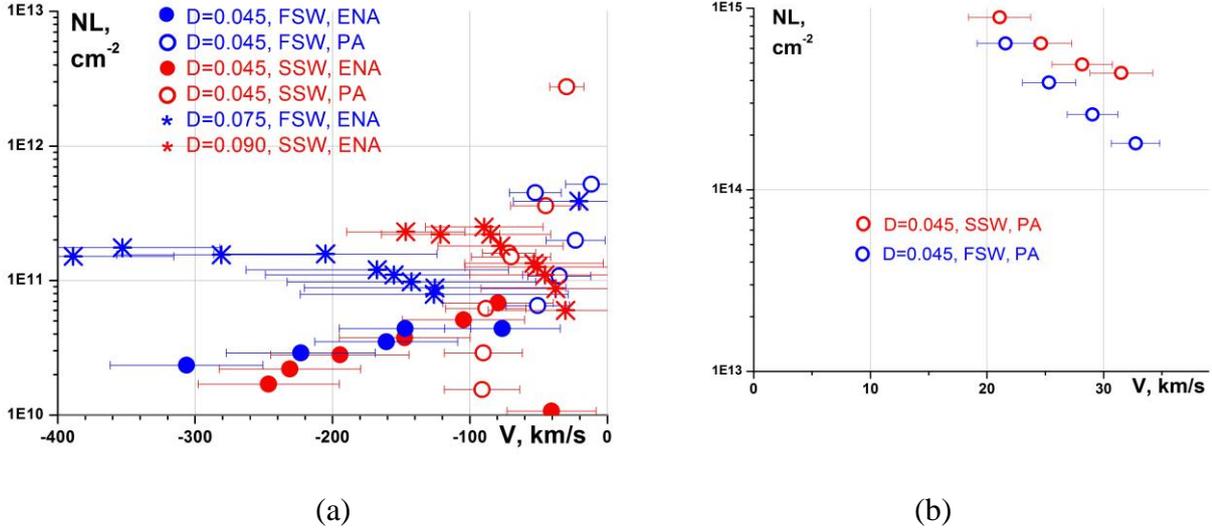

(a)          (b)

**Figure 15.** Column density NL of hydrogen atoms versus an average density weighted bulk velocity $\tilde{V}_z$, each calculated along the same lines-of-sight evenly spaced by $1R_p$ across the planet-star line. Figure is split into negative (a) and positive (b) velocity domains. The color of data points, blue and red, corresponds to the fast (FSW) and slow (SSW) stellar wind, respectively. Circles and asterisks are used to distinguish between the *"captured by the star"* (D=0.045AU with FSW and SSW) and *"blown by the wind"* (D=0.075AU with FSW and at D=0.09AU with SSW) regimes respectively. Filled and empty circles mark NL values caused by ENAs and planetary atoms, respectively. Bars indicate density weighted thermal dispersion.

The precise calculation of Lyα absorption by the ENA cloud simulated with our model is not the goal of the present paper. This subject deserves a separate study which is planned to be done in combination with a Monte-Carlo model (*Holmstrom et al. 2008, Kislyakova et al. 2014a,b*) to reproduce better the complex evolution of the planetary atoms beyond the ionopause and their interaction with the protons of SW. At the same time, a rough estimation of possible effect in the transiting spectra of different PW and SW interaction regimes investigated in the present paper, could be done by comparison of column density integrals for the neutral hydrogen (i.e., planetary atoms $n_H^{pw}$ and ENAs $n_H^{sw}$), calculated along the line-of-sight (i.e. the planet-star line) at different locations around the planet. The outcome of this study is summarized in *Figure 15*.

Note, that in the case of the *"captured by the star"* regime, we distinguish between the area of the planetary material stream flowing towards the star, which will contribute the red-sifted absorption (*Figure 15b*), and the entire peripheral area outside of the stream (*Figure 15a*). By this, due to the interstellar medium absorption, it is unlikely that Lyα features caused by hydrogen atoms in the stream of planetary material falling towards the star would be visible. At the same time the heavier species which will be dragged by the stream might still produce measurable effects. Because of an insignificant variation of the velocity across the stream and its almost linear increase in the longitudinal direction (e.g., see *Figure 12*), the column density inside the stream $NL(Z) = \int_{Z}^{Z+dz} n_a d\tilde{z}$ is integrated in steps $dz=1R_p$. Thus, each positive velocity value on the horizontal axis in *Figure 15b* corresponds to a certain position $0 < Z < 10R_p$ along the stream.

For the area outside the planetary material stream, as well as in the case of the *"blown by the wind"* regime, we calculate for a number of lines-of-sight evenly spaced with the step $1R_p$ (up to $10R_p$) across the planet-star line, the column density $NL(r) = \int_{-10R_p}^{10R_p} n_a d\tilde{z}$ and a density-weighted average velocity $\tilde{V}_z(r) = \int V_z n_a d\tilde{z} / \int n_a d\tilde{z}$, taking into account the thermal velocity $V_T = \sqrt{k\tilde{T}/m}$ dispersion in terms of the density-weighted temperature $\tilde{T}(r) = \int T_a n_a d\tilde{z} / \int n_a d\tilde{z}$. Instead of $NL(r)$ we plot dependence $NL(\tilde{V}_z)$ to indicate typical velocities at which Lyα is absorbed. Thus, each negative velocity value on the horizontal axis in *Figure 15a* corresponds to the velocity $\tilde{V}_z(r)$ calculated at particular radial distance r<10Rp. Note, that the column density of the streaming towards the star PW is two orders of magnitude higher than the analogous values outside the stream as well as outside the plasmasphere formed in the case of *"blown by the wind"* regime.

The overall conclusion, that can be made with the obtained modelling results is, that in the *"captured by the star"* regime, which might take place on the real HD209458b, the tidally pulled stream of the escaping planetary material reaches sufficiently high velocities and may be considered as a potential driver which could drag species heavier than hydrogen to explain the observed absorption in corresponding lines. However, the Doppler-shifted by more than 100 km/s absorption, especially in blue wings of Lyα line, caused by ENAs cloud, requires much more planetary atomic hydrogen, reaching the SW, than that seen in the simulations. In that respect, it should be mentioned that there are several factors, not included in our model, which might affect photo-absorption and photo-ionization and therefore influence the amount of atomic hydrogen in the PW. These are for example the presences of helium and metals in the planetary atmosphere. The question of ionization balance in a complex PW which includes, besides of hydrogen, also the heavier species, requires further and more detailed study in order to access the importance of ENAs in absorption spectroscopy in the case of HD209458b. Note, that as it has been demonstrated in *Ben-Jaffel and Sona Hosseini (2010)* and *Kislyakova et al. (2014a)*, the effects of spectral line broadening may produce a significant contribution in the formation of the observed transiting spectra of HD209458b.

As to the explanation of observations with particle models (*Bourrier & Lecavelier des Etangs 2013, Bourrier et al. 2015*), which with the properly chosen parameters of the surrounding material enable the reproducing of observational effects, the next important step would be to physically justify these parameters under the conditions of a particular planet. In that respect, the results of our self-consistent hydrodynamic simulations imply that the parameters of PW plasma used particle models need more rigorous validation.

### Acknowledgements

This work was supported by grants № 14-29-06036, 16-52-14006 of the Russian Fund of Basic Research, RAS SB research program (project II.10.1.4, 01201374303), as well as by the projects S11606-N16 and S11607-N16 of the Austrian Science Foundation (FWF). MLK also acknowledges the FWF projects I2939-N27, P25587-N27, P25640-N27 and Leverhulme Trust grant IN-2014-016. Parallel computing simulations, key for this study, have been performed at the Supercomputing Center of the Lomonosov Moscow State University, at SB RAS Siberian Supercomputer Center, and at Computation Center of Novosibirsk State University.